\documentclass[10pt,a4paper]{article}
\usepackage{graphpap,epsfig,amssymb,graphicx,textcomp,amsmath,xcolor}
\usepackage[export]{adjustbox}
\usepackage{hyperref}

\begin{document}

\begin{center}

{\LARGE {\bf What will be the Euclidean dimension of an Ising ferromagnetic cubic shell?}}

\vskip 0.5cm

{\large {\it Ishita Tikader$^1$ and Muktish Acharyya$^{2,*}$}} \\
{\it Department of Physics, Presidency University,}\\
{\it 86/1 College Street, Kolkata-700073, INDIA}\\
{$^1$E-mail:ishita.rs@presiuniv.ac.in}\\
{$^2$E-mail:muktish.physics@presiuniv.ac.in}

\end{center}

\vskip 1cm

\noindent {\bf Abstract:} The equilibrium and nonequilibrium properties of an Ising ferromagnetic cubic shell have been extensively studied by Monte Carlo simulation using Metropolis single spin flip algorithm. Although, geometrically the Euclidean dimension of the cubical shell is three, interestingly, the Ising ferromagnetic cubic shell undergoes ferromagnetic phase transition at a temperature which is very close to that for two-dimensional Ising ferromagnet. Surprisingly, the Ising ferromagnetic cubic shell shows
a strange (neither exponential nor stretched exponential) kind of relaxation 
behaviour, instead of exponential relaxation as usually observed in the two dimensional Ising ferromagnet. The metastable lifetime of a ferromagnetic Ising cubical shell is studied as a function of the applied magnetic field. Here also, the cubic shell behaves more likely a two-dimensional object as found from statistical analysis and comparison with Becker-D\"oring prediction of classical nucleation theory. 

\vskip 3cm

\noindent {\bf Keywords: Euclidean dimension; Ising ferromagnet; Monte Carlo simulation; Metropolis algorithm; Critical point; Relaxation; Metastable behaviour}

\vskip 2cm
\noindent {\bf PACS Nos:} 05.10.Ln; 05.70.Fh; 75.10.Hk; 75.60.Jk; 64.60.My

\vskip 2cm

\noindent $^*$ Corresponding author
\newpage

\section{\bf {Introduction:}}
The equilibrium phase transition of a ferromagnetic sample is well studied\cite{stanley}. Since, the Ising ferromagnet is a prototype of ferromagnet, the equilibrium ferro-para phase transition is well studied by Monte Carlo simulation. In two-dimensions, the model (with nearest neibour interaction) is exactly solvable. In three-dimensions, it hase been studied by
Monte Carlo simulations.
The relaxational behaviours of a ferromagnetic sample is also studied widely. The Ising ferromagnet has been found to show
the exponential relaxation both in two and in three dimensions. 

The ferromagnetic samples show the metastable behaviour. This metastability is a subject of modern research in the context
of preparing the magnetic storage devices.
In our everyday life, magnetic thin films are widely used in a range of devices due to their enormous technological applications. Magnetic data storages \cite{storage} are essential and useful devices that play a crucial role in storing information in tiny magnetic grains. The switching of magnetisations by an externally applied magnetic field determines the dynamical behaviours of
such storage devices. In fact, the devices for fast recording require rapid switching of magnetisation. For fast recording or rapid accessing the data the magnetic systems are expected to respond quickly to the variations of the external magnetic field. Not only this quick response is sufficient for good storage devices. The magnetic systems should possess adequate stability
against any kind of noise (thermal or any non-thermal)\cite{vogel}.

The subject of switching of magnetisation by externally applied weak magnetic field is important to study. Theoretically or historically
the problem was investigated  by Becker and D\"oring (in 1935)\cite{becker,moumita-review} 
through their celebrated classical nucleation theory. The reversal time (or nucleation time) has been found phenomenologically
as a function of temperature and applied magnetic field in general d-dimensions. This picture provides the metastable lifetime
and its dependences on the temperature and applied magnetic field. The rate of nucleation of crystalline solids in a solid-melt system was explored in an important historical study\cite{grant}. 

A considerable amount of theoretical and experimental works has been done over the last ninety years after Becker-D\"oring
theory. The simple and prototypical model of ferromagnet, the Ising ferromagnet, has been chosen to study the reversal phenomena. The relaxation of Ising ferromagnet after a sudden reversal of applied magnetic field is also studied\cite{binder1}. The dynamics of the domains or droplets of magnetic films with perpendicular anisotropy have been reported\cite{nowak}. 
Good agreement between the numerical results and the theoretical predictions of Becker-D\"{o}ring has been reported\cite{acharyya1} through extensive simulational investigations on nucleation in different dimensions  using heat-bath dynamics. The prediction of this phenomenological theory was verified by the Monte Carlo simulation where the growth of droplets has been studied as phase ordering kinetics\cite{puri1}. The rates of growth and decay of the droplets of different sizes have been studied \cite{vehkamaki} as functions of external field and temperature. The heat-assisted magnetisation reversal in ultrathin films for ultra-high-density information recording has been investigated\cite{deskins}. The extensivity property of
the metastable lifetime has also been reported\cite{rikvold}. Very recently, the general S-spin Ising model has been used to 
study\cite{mapre21} the metastable behaviour and compared the results with Becker-D\"oring analysis.
However, the literature of metastable behaviours in ferromagnetic system is lacking by studies of the role of geometry.
Recently, the metastablity can be affected by the geometry of the system. The relaxation behaviour of an Ising strips (of equal area)
has been studied \cite{iom} by Monte Carlo simulation. 

 {\it What will be the Euclidean dimension of an Ising ferromagnetic cubic shell?} Geometrically, it is a three dimensional
(Euclidean) object. But will it behave like a three dimensional object in the context of ferromagnetic behaviour ? To address this question, we report, in this paper, the Monte Carlo results of the equilibrium behaviour, relaxational behaviour and the metastable behaviour
of an Ising ferromagnetic cubic shell. Will it show its tri-dimensionality in the context of ferromagnetic phase transition ? Will it exhibit its
tri-dimensionality in the context of magnetic relaxation behaviour? Will it
reveal its tri-dimensionality in the context of growth of droplets in the reversal of magnetisation? All these questions are addressed 
(through extensive Monte Carlo investigation) in this
article. The manuscript is organized as follows: the model of the cubical shell of Ising ferromagnet along with
the simulational scheme are mentioned in the
next section, the numerical results are provided in section-3, the paper ends with summary in section-4.

\section {\bf {Model and Simulation:}}
This work focuses on the classical Ising model with nearest neighbour ferromagnetic interactions. In the presence of an external magnetic field, the Hamiltonian reads as,
\begin{equation}
\label{hamiltonian}
H=-J\sum_{<i,j>}S_iS_j - h_{ext} \sum_{i}S_i
\end{equation}
where the Ising spin variables $S_i$ are allowed to have values $S_i \in \{-1, +1\}$; $\forall~i$. The first term arises due to the interactions between nearest neighbouring (NN) spins.  Here, $J(>0)$, stands for the uniform ferromagnetic coupling.  The second term, the Zeeman term, in the Hamiltonian represents the interaction energy with external magnetic field $h_{ext}$.

The Monte Carlo simulations have been executed on a cubic shell of size $L$, having $N_s= L^3 - (L-2)^3 = 6L^2-12L+8$ number of lattice sites with open boundary conditions in all three directions. \textcolor{blue}{The reason, for the implementation of open boundary 
condition here, is as follows: the spin of each surface will interact with the neighbouring spins of that surface only except from the spins at the edges. The spins at the edges will interact with neighbouring spins of two adjuscent surfaces. The periodic boundary condition, if implemented, would provide additional interaction for the spins located on any surface of the cubical 
shell.}

\begin{figure}[h!tpb]
 \centering
  \includegraphics[angle=0, width=1.00\textwidth]{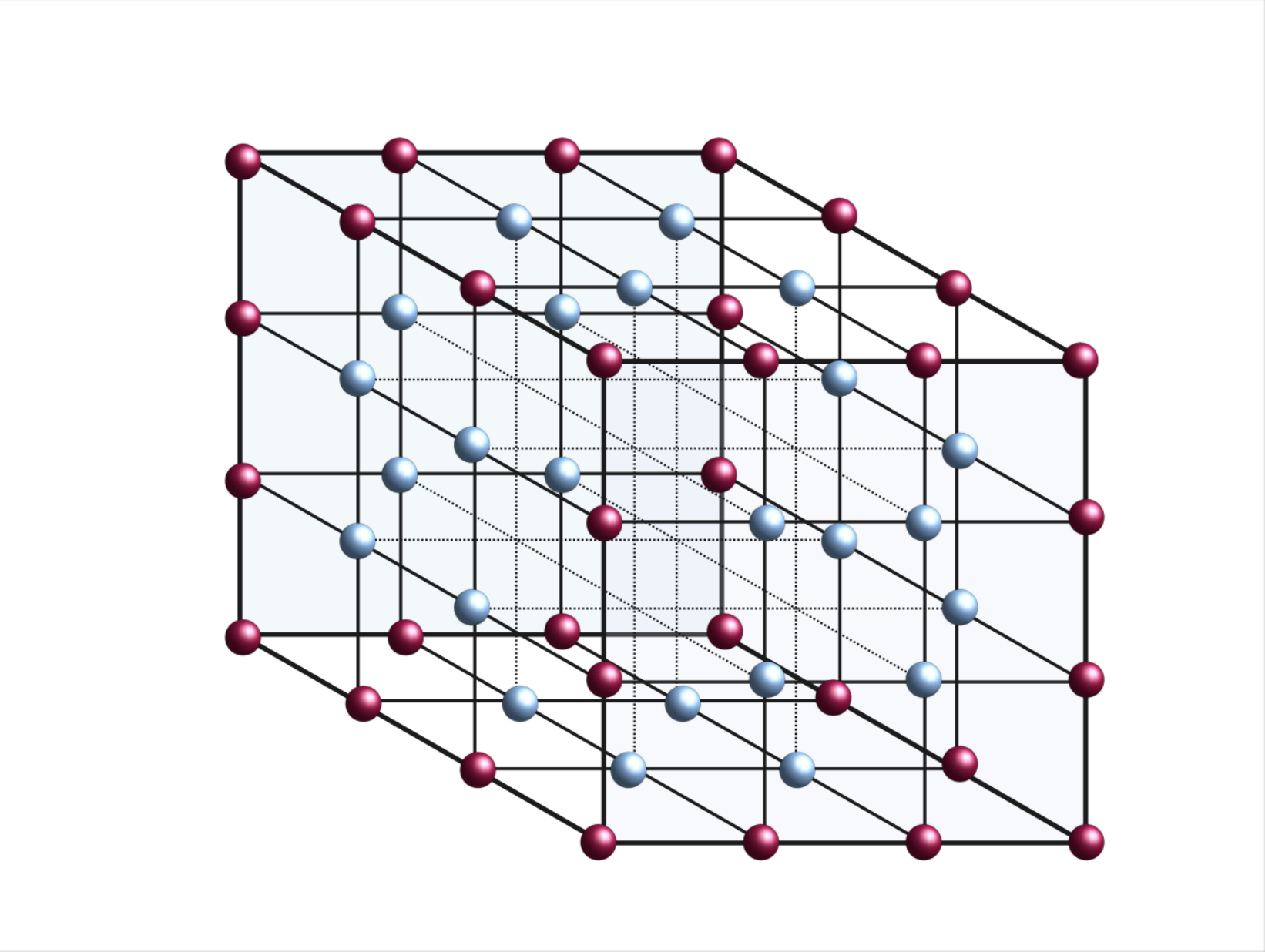}
  \caption{A schematic demonstration of a cubic shell of size $L$, having lattice sites $N_s = L^3 - (L-2)^3$. The red ball stands for the site on the edge and the blue one represents the interior point on the surface.}
  \label{fig:spinstructure}
\end{figure}

As a simple demonstration, a schematic diagram of a cubic shell of $L=4$ is shown in  figure \ref{fig:spinstructure}. The blue coloured balls denote the spins on the suface of the cubic shell lattice, while red coloured balls indicate the spins on the edge or boundary of the lattice. Inside the cube, there is no
such magnetic spins. At any specified temperature ($T$),
the probability of spin-flip at any site $i$ (randomly chosen)  is determined by Metropolis algorithm\cite{binder}.
\begin{subequations}
    \begin{equation}\label{prob}
       P(S_i \rightarrow  -S_i) = Min \Biggl[ 1, \exp \biggl(-\frac{\Delta E_i}{k_BT} \biggr) \Biggr]  
    \end{equation}
\rm{where $\Delta E_i$ is the change in energy (measured in unit of $J$) due to spin-flip at $i^{th}$ site.}
    \begin{equation}\label{local_energy}
      \Delta E_i =  2S_i \Biggl[ J\sum_{NN}S_j + h_{ext} \Biggr]
    \end{equation}
\end{subequations}
$k_B$ is the Boltzmann constant and $T$ is the temperature of the system measured in the unit of $J/k_B$. For simplicity, we have assumed $J =1$ and $k = 1$ throughout our simulational study. The $N_s$ number of such random updates is defined as one
Monte Carlo Step per Site (MCSS).\par

At each MCSS, the instantaneous magnetisation is formulated as follows,
\begin{equation}\label{m(t)}
   \Tilde{m}(t)= \frac{1}{N_s}\sum_{i} S{_{i}}. 
\end{equation}
To determine the pseudo-critical point of the cubic shell with Ising spins the equilibrium magnetisation (in the absence of external field, i.e., $h=0$) is measured as $M=\langle \Tilde{m} \rangle$. We have also determined susceptibility as, $\chi = \frac{N_s}{k_BT}(\langle \Tilde{m}^2 \rangle - {\langle \Tilde{m} \rangle}^2)$. Here $\langle .. \rangle$ denotes the time average of thermodynamic quantities, approximately equal to the ensemble average in the presumed ergodic limit.

\section {\bf {Results:}}
Before discussing our simulational results, we address a simple but intriguing question. What is the Euclidean dimension of Ising ferromagnetic cubic shell? To answer this simple question we have studied the equilibrium behaviour or ferromagnetic phase transition as well as transient behaviour e.g. magnetic relaxation and decay of the metastable state in a cubic shell-structured lattice with Ising spins. 
\subsection{Equilibrium behaviour:}
To study the equilibrium ferromagnetic phase transition, we have slowly cooled down the system from a high temperature (paramagnetic or disordered phase i.e., $M=0$) random spin configuration.  Near the critical temperature, the magnetisation $M$ of the system starts to have nonzero value i.e., $M \neq 0 $ at $T<T_c$; while the susceptibility ($\chi$) exhibits a sharp peak (eventually diverges in thermodynamic limit, $L \to \infty$).  The pseudo-critical temperature $T_c^p$ for the finite-sized system is effectively determined by the maxima of susceptibility. 

\begin{figure}[h!tpb]
\begin{center}
  \includegraphics[angle=0, width=1.00\textwidth]{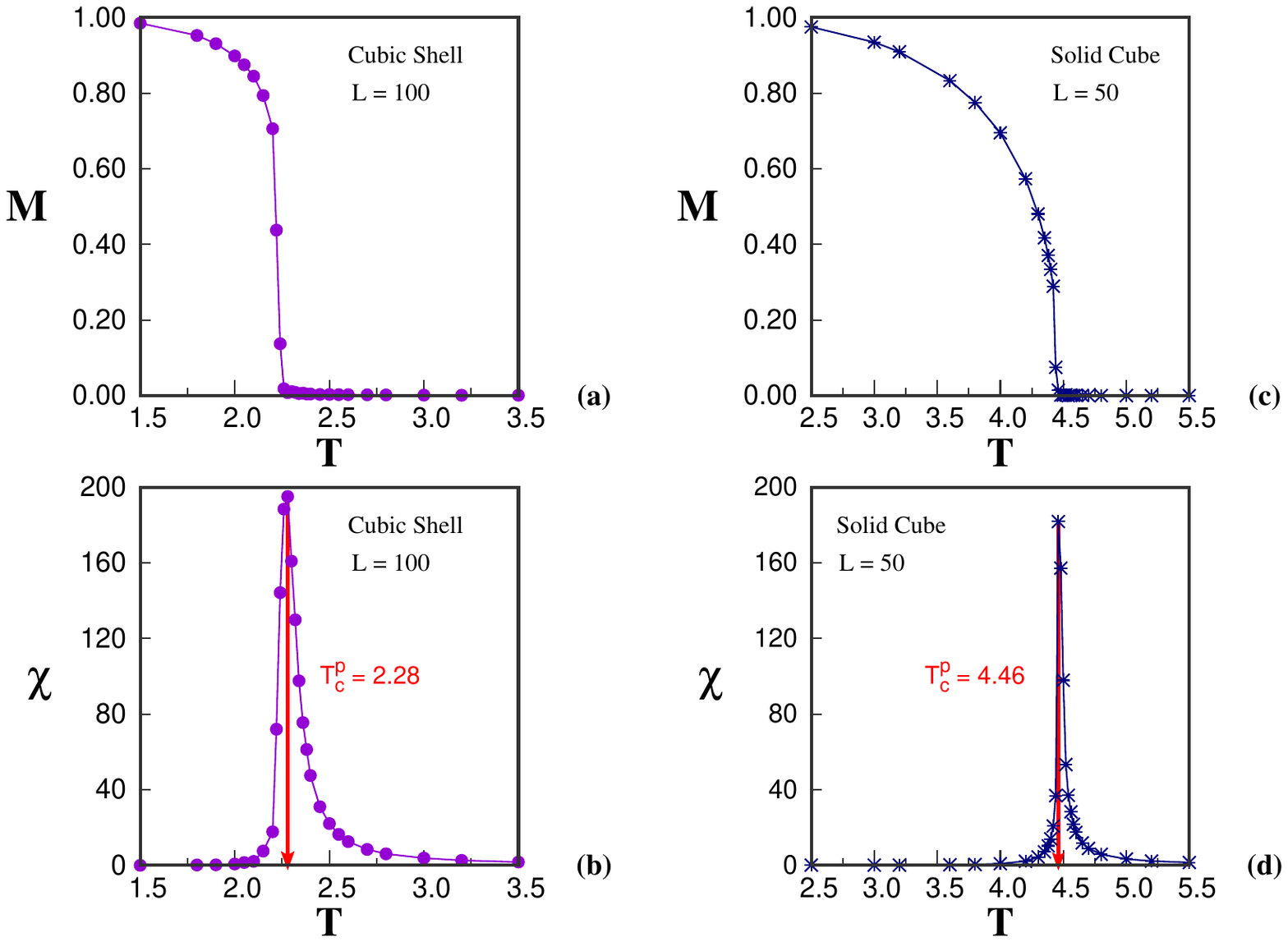}
  \caption{Comparative study of thermodynamic quantities (equilibrium magnetisation $M$ and susceptibility $\chi$) as the functions of temperature ($T$). The figures in the left panel show the results for the Ising ferromagnetic cubic shell of size $L=100$; the temperature dependence of (a) the magnetisation ($M$) and  (b) the susceptibility ($\chi $) represented in violet solid circle ($\bullet$). The figures in the right panel show the results for the solid cube of size $L=50$; the temperature dependence of (c) the magnetisation ($M$) and (d) the susceptibility ($ \chi $) represented in blue asterisk ($\boldsymbol{\ast}$).}
  \label{fig:Criticalpoint}
\end{center} 
\end{figure}

In this study, we have investigated the temperature dependence of magnetisation and susceptibility in the Ising ferromagnetic system modelled on a \textcolor{blue}{cubic shell of size $L=100$. The thermodynamic quantities ($M~\rm and ~ \chi$) are recorded up to 75000 MCSS time steps, out of which the initial 65000 MCSS time-steps are discarded to achieve equilibrium. These statistical quantities ($M ~ \rm and ~ \chi$) are determined by taking the average over the last $10^4$ MCSS. Furthermore, we have calculated the sample average of $M ~\rm and ~ \chi$ over 250 random samples. We have cooled down the system with small step size 
($\delta T$) of the temperature. The step size is chosen $\Delta T = 0.02$ near the transition temperature. However, larger 
step size ($\Delta T=0.1$) has been chosen for the temperature much away from the critical temperature.} To determine the transition point, the temperature dependences of both magnetisation $M$ and susceptibility $\chi$ are studied and graphically depicted in figure \ref{fig:Criticalpoint}(a) and \ref{fig:Criticalpoint}(b). The average magnetisation ($M$) has been found to grow continuously from zero to nonzero, as the system cools down and the  susceptibility ($\chi$) has been found to get peaked at a particular temperature $T=2.28$. This indicates the continuous ferro-para phase transition of an Ising ferromagnetic cubic shell. This transition point is defined as the pseudo-critical temperature of the Ising ferromagnetic cubic shell, i.e., $T_c^p = 2.28$ (in unit of $J/k_B$) for the finite-sized ($L$ finite) cubic shell of $L=100$. In the thermodynamic limit ($L \rightarrow \infty$) the value of the critical temperature $T_c$ for Ising cubic shell should be slightly different from $T_c^p$. However, it is worth mentioning that, the value of $T_c^p$ for the cubic shell is remarkably close to the Onsager value of $T_c=2.269...$\cite{Onsager} (in unit of $J/k_B$) for the two-dimensional Ising system. In the context of ferromagnetic phase transition, the Ising ferromagnetic cubic shell behaves here like a two-dimensional 
object. This result is not at all surprising. Since the Ising ferromagnetic cubic shell is a structure where each spin (except the
spins at eight corners) has four nearest neighbours, its behaviour regarding the equilibrium ferro-para phase transition resembles that of a two-dimensional system.  

We have also observed the phase transition in the solid cubic Ising ferromagnetic system of lattice size $L=50$ with open boundary condition, analysing the variation of average magnetisation and susceptibility with temperature. The pseudo-critical temperature for solid cubic lattice is marked to be $T_c^p = 4.46$ (figure \ref{fig:Criticalpoint} (c) and (d)), which is close to the Monte Carlo estimate of the critical temperature of a three dimensional system $T_c = 4.511$ \cite{Ito-Suzuki}. The comparative study of the transition point between a cubic shell and a solid cube is demonstrated in figure \ref{fig:Criticalpoint}. So, we have come to the conclusion that the ferro-para phase transition in Ising ferromagnetic cubic shell matches with that for a flat two-dimensional system, even though geometrically both cubic shell and solid cube are three-dimensional objects. {\it Should we expect to have the
two dimensional behaviour of an Ising ferromagnetic cubic shell in the context of relaxational behaviour?} To find the answer
to this question, we have studied the relaxation behaviour of an Ising ferromagnetic cubic shell and compared the results with that for a solid cube. 
 
Since we are eager to examine transient behaviours like magnetic relaxation, and metastable dynamics of Ising ferromagnetic cubic shell along with a comparative study between cubic shell and solid cube,  the equivalent thermal condition should be maintained in both cases. For that reason, we have chosen the temperature of the respective systems  with same multiplicative factor of respective critical temperatures ($T_c$). Precisely, if the temperature for the Ising ferromagnetic cubic shell is $T=f \times T_c^{Shell}$,
the same thermal condition will be provided for solid cube at temperature $T=f \times T_c^{solid}$. In the next subsection, we 
will discuss the magnetic relaxation behaviour of both Ising ferromagnetic cubic shell and solid cube.

\subsection{Relaxation behaviour:} The ferromagnetic model system exhibits well studied relaxation behaviours. If the system is initially configured with all spins up ($S_i = +1 ~ \forall ~ i$) at any finite temperature T, it will eventually relax and reach its equilibrium state corresponding to that temperature ($T$). Here, we have studied such kind of relaxation behaviour
of an Ising ferromagnetic cubic shell.
Consider an Ising ferromagnetic cubic shell (of size $L=50$), initially $S_i=+1 ~\forall ~i$. 
Certainly, this is not the equilibrium state of the system at any finite nonzero temperature $T$.
We are
 trying to achieve its equilibrium state for a temperature slightly above the corresponding critical temperature. Now let the system relax in the absence of an external magnetic field ($h_{ext} = 0$). The magnetisation of the system $m(t)$ starts to decay with time and eventually vanishes. This relaxation process is not instantaneous; it takes a certain time duration and finally, the system achieves the stable equilibrium state of $m=0$. 

\begin{figure}
\centering
 \includegraphics[angle=0, width=1.00\textwidth]{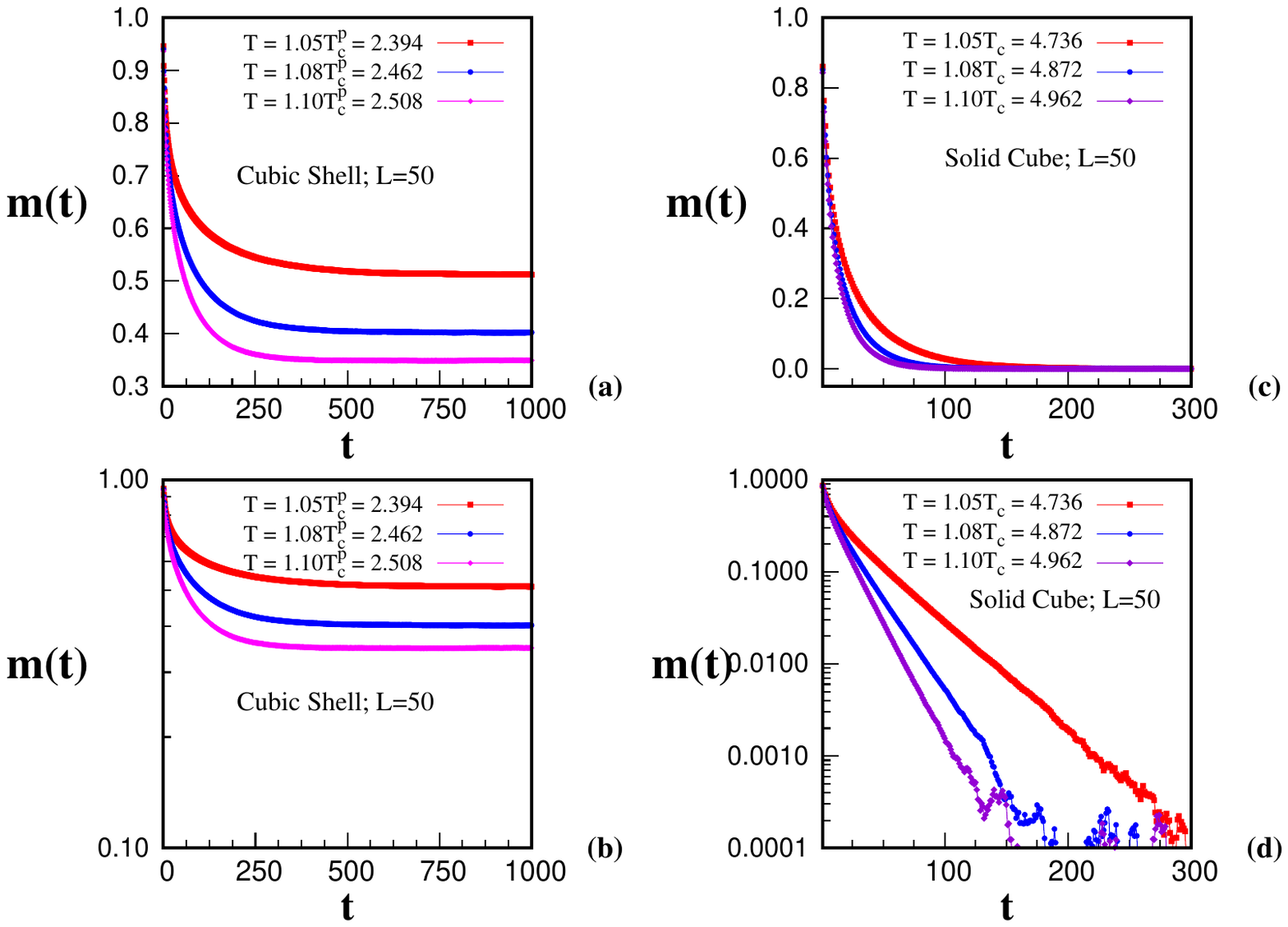}
 \caption{Relaxation of magnetisation. The time evolution of magnetisation $m(t)$ in the Ising ferromagnetic cubic shell ($L=50$) plotted in (a) linear scale and (b) semi-log scale. Results obtained for for three different temperatures in the paramagnetic phase i.e. $T=1.05 T_c^p, 1.08T_c^p$ and $1.10T_c^p$. Here $T_c^p = 2.28$ for cubic shell lattice. Similarly the time variation of magnetisation $m(t)$ in solid cubic lattice of size $L=50$ plotted in (c) linear and (d) semi-log scale. Here also we have noted the results for three temperatures $T=1.05T_c, 1.08T_c$ and $1.10T_c$. $T_c=4.511$ for 3D Ising system.}
 \label{fig:relaxation}
\end{figure}

We have investigated the magnetic relaxation in the Ising ferromagnetic cubic shell for three different temperatures (fixed at $T=1.05T_c^p, ~ 1.08T_c^p$ and $ 1.10T_c^p$), as depicted in figure \ref{fig:relaxation} (a) and (b). The instantaneous magnetisation $m(t)$ is determined by taking average over 10000 random samples to ensure the smooth variation with time. We have studied the  instantaneous  magnetisation ( sample averaged) as a function of time ($t$) and plotted in both linear and semi-logarithmic scale for cubical shell as well as a solid cubic lattice of size $L=50$ with open boundary condition. Figure \ref{fig:relaxation} (a) and (b) illustrate the time evolution of magnetisation for cubic shell  in linear and semi-log scale respectively, while figure \ref{fig:relaxation} (c) and (d) represent those for the solid cube at three different temperatures. The theoretical \cite{suzuki} study as well as Monte Carlo simulation  predicted\cite{iom} the exponential nature of magnetic relaxation both in two and three-dimensional Ising ferromagnetic systems near critical point. The variation of magnetisation with time in semi-log scale (Fig-\ref{fig:relaxation}(d)) gives a straight line for solid cubic lattice,
revealing the {\it exponential} relaxation. Whereas our results for cubic shell exhibit a non-exponential relaxation in the semi-log plot 
(not linear) of $m(t)$ ~versus~$t$ (Fig-\ref{fig:relaxation}(b)). {\it Here is the surprise !!} We have seen (in the previous subsection) that the equilibrium ferro-para phase transition of an Ising ferromagnetic cubic shell compels us to believe that it behaves magnetically like a two-dimensional system. But the two dimensional Ising ferromagnet shows the exponential behaviours as contradicted by our
present result of non-exponential relaxation of an Ising ferromagnetic cubic shell. \textcolor{blue}{Usualy, exponential 
relaxational is observed in two-dimensional Ising ferromagnet and that is theoretically understood in the linearized mean field equation of Glauber kinetic Ising ferromagnet\cite{suzuki}. Those results are obtained for the compact solid systems. However, the studies on relaxational behaviours for the hollow systems (or fractal like systems) are missing in the literature. 
The systems (not simply connected) with large cavity may show different relaxtion behaviours. It is not yet understood the basic reason of having non-exponential
relaxation in the case of cubic shell of Ising ferromagnet. The results of our present study may trigger further research to
investigate the basic reasons of magnetic relaxation related to the geometry (not simply connected)  of the system. It may 
be worth mentioning here that the non-exponential magnetic relaxation has recently been observed experimentally in magnetic nanoparticles \cite{gresits}. The single molecular magnet of manganese also showed non-exponential magnetic relaxation\cite{yamaguchi}. }

\begin{figure}
\centering
 \includegraphics[angle=0, width=1.00\textwidth]{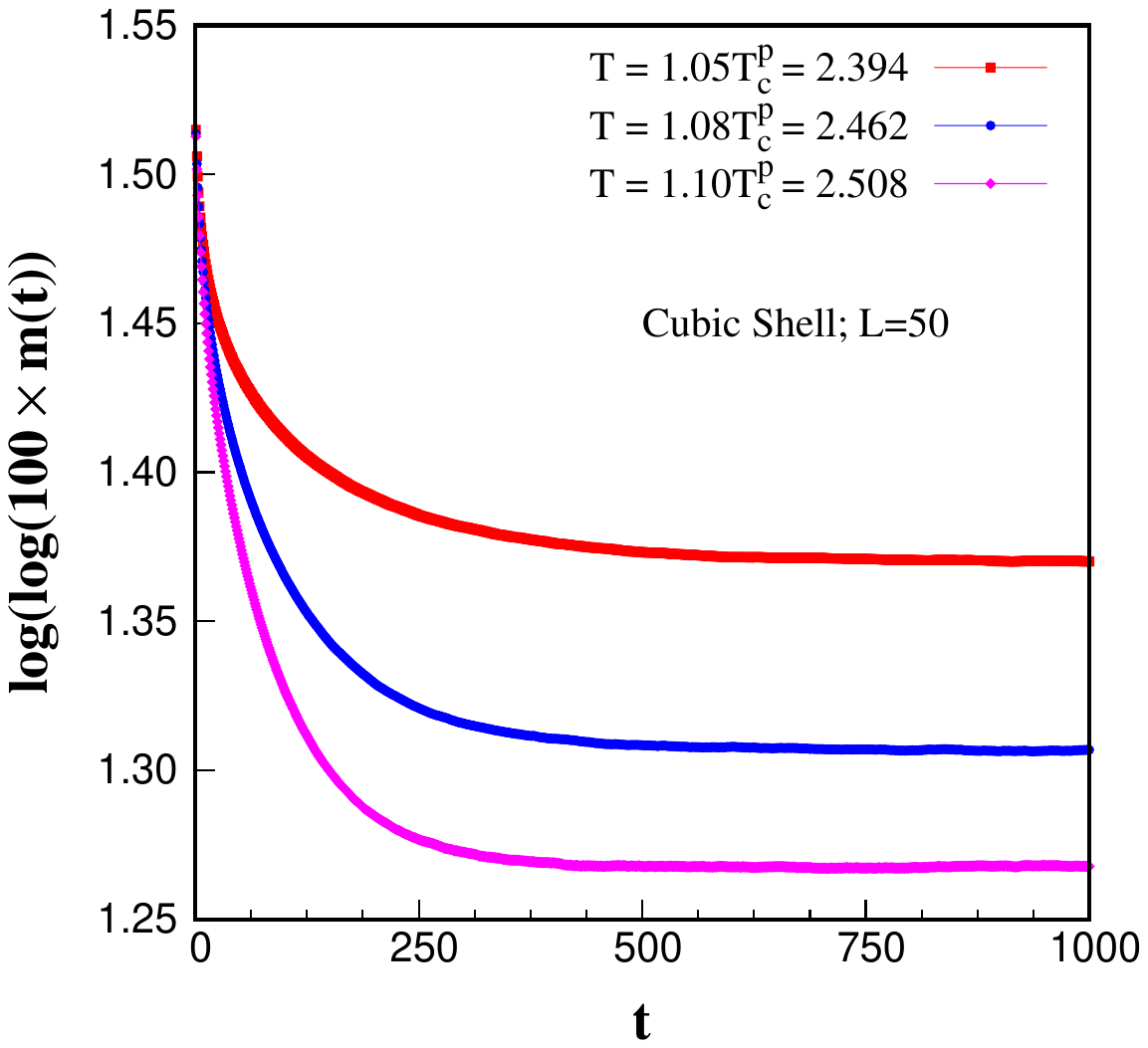}
 \caption{$\log(\log(100\times m(t)))$ is plotted against time (in unit of MCSS) for three different temperatures in the paramagnetic phase i.e. $T=1.05 T_c^p, 1.08T_c^p$ and $1.10T_c^p$. Here pseudo-critical temperature of Ising ferromagnetic cubic shell $T_c^p = 2.28$.}
 \label{fig:loglogm}
\end{figure}

\begin{figure}
\centering
 \includegraphics[angle=0, width=1.00\textwidth]{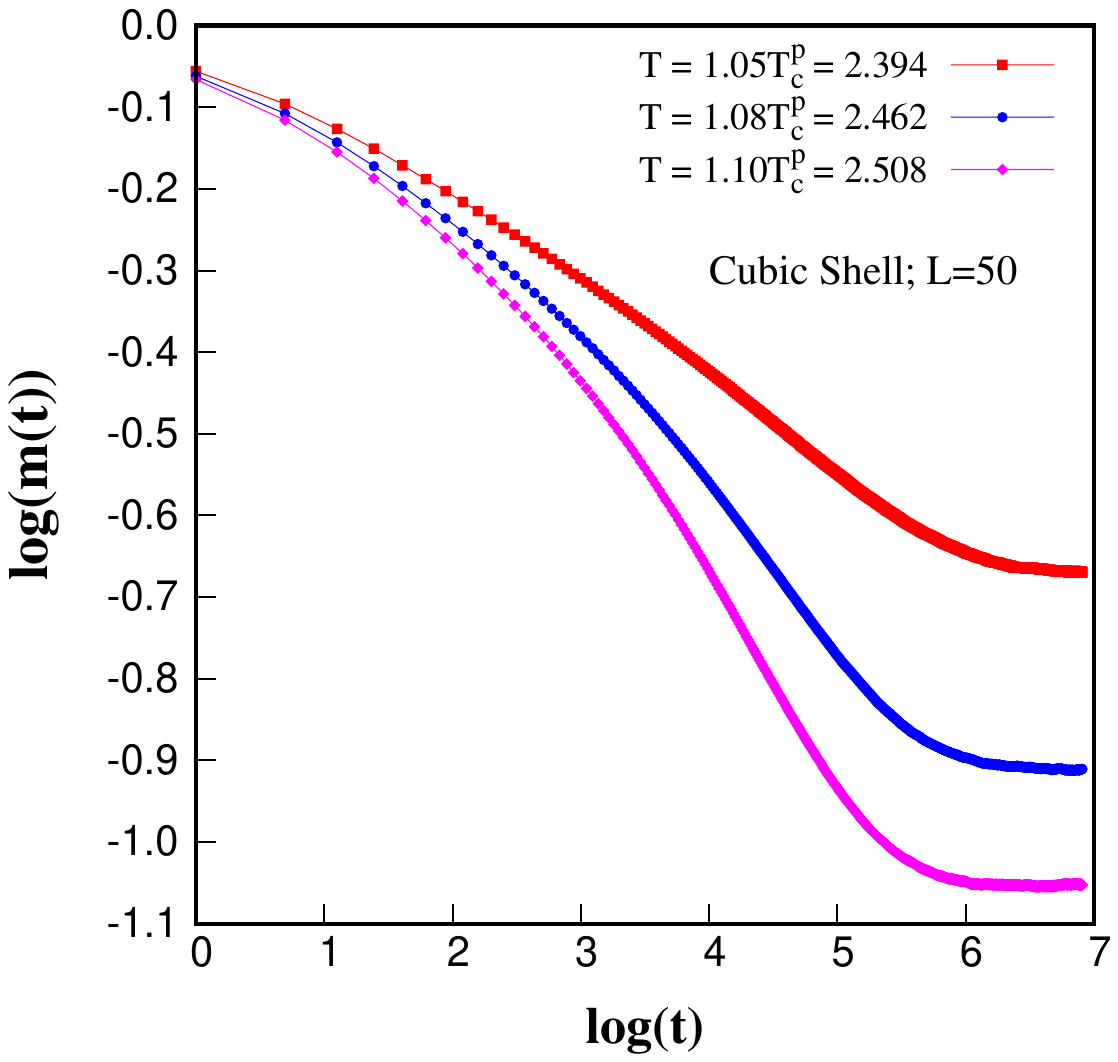}
 \caption{The logarithmic value of magnetisation $\log(m(t))$ is plotted as a fuction of $\log(t)$ for three different temperatures in the paramagnetic phase i.e. $T=1.05 T_c^p, 1.08T_c^p$ and $1.10T_c^p$. Here pseudo-critical temperature of Ising ferromagnetic cubic shell $T_c^p = 2.28$. }
 \label{fig:logm_logt}
\end{figure}

So, what kind of relaxation behaviour is expected from such a cubic shell ? Is it stretched exponential ? Is it power law type ?
We have examined the variation of $\log(\log(m(t))$ with time to ascertain whether the cubic shell undergoes stretched exponential relaxation or not. Fig-\ref{fig:loglogm} shows the plot of $\log(\log(100 \times m(t))$ as a function of time ($t$). The results indicate that magnetic relaxation of the Ising ferromagnetic cubical shell does not even show the stretched exponential behaviour. The logarithmic value of magnetisation $\log(m(t))$ is also studied as a function of $\log(t)$ and shown in  Fig-\ref{fig:logm_logt}. This graph  also fails to indicate any power-law type of relaxation behaviour. May we conclude from the obtained results of the non-exponential relaxation behaviour that  the Ising ferromagnetic cubic shell does not even behave like a geometrically two-dimensional object, as predicted from its equilibrium ferro-para phase transition (section-3.1)?
{\it So, what will
be the Euclidean dimension of an Ising ferromagnetic cubic shell?}

The strange relaxation behaviours of the cubic shell prompted us to proceed further to investigate its metastable behaviour and compare that with the
prediction of classical nucleation theory. The classical nucleation theory (Becker-D\"oring theory) systematically and critically provides the role
of Euclidean dimensions of the ferromagnetic system through its (dimensional) dependence reflected in the magnetisation reversal time.
The next subsection is devoted to the Monte Carlo study of metastable behaviours of an Ising ferromagnetic cubic shell and
that of a solid cubic Ising ferromagnet.

\subsection{Metastable behaviour:} The ferromagnetic system shows the metastable behaviours. The idea is to start with 
all spin up configuration (metastable state) in the presence of magnetic field applied in opposite direction. Now evaluate the time required by 
the system (the metastable lifetime) to achieve the stable state (a state of magnetisation in the opposite direction of
initial magnetisation). This time is metastable lifetime which depends critically on the Euclidean dimension of the 
system.

\begin{figure}
    \centering
    \includegraphics[angle=0, width=1.00\textwidth]{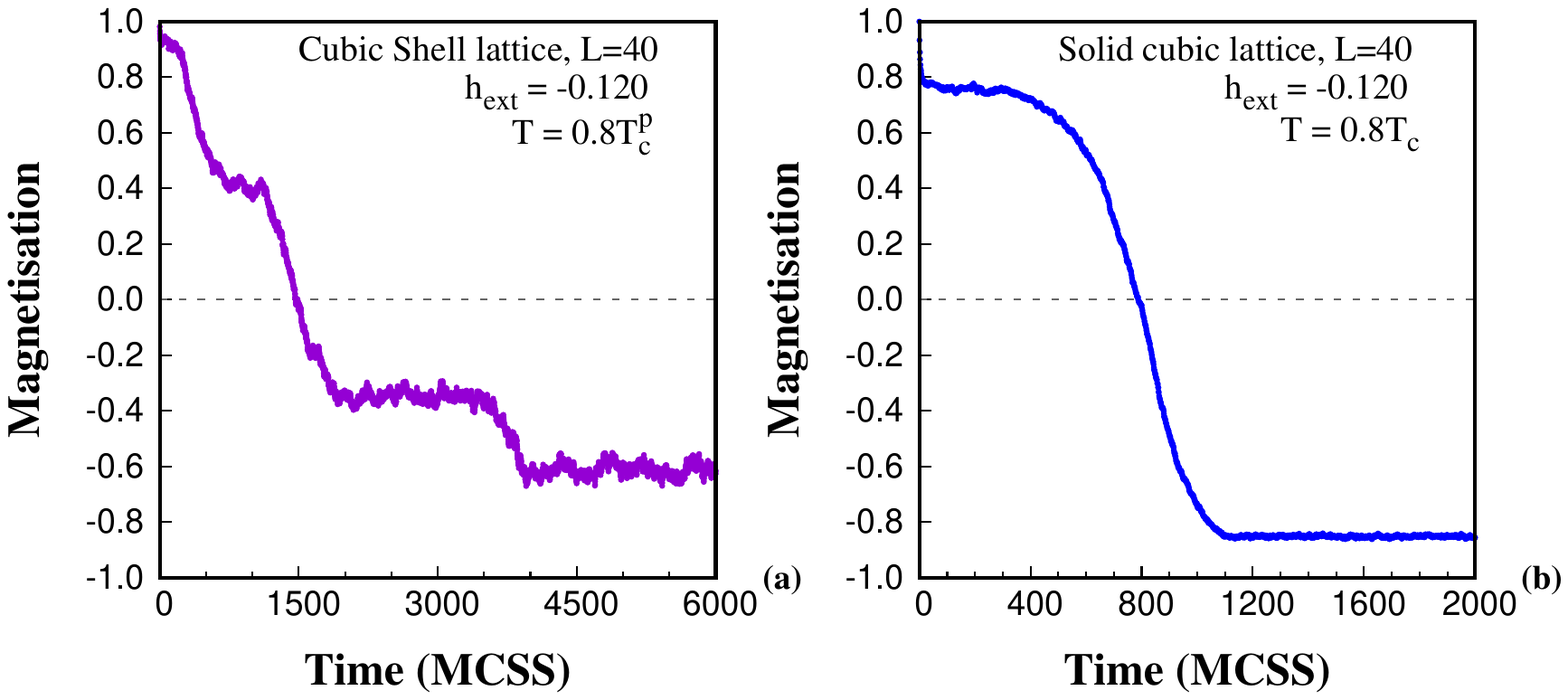}
    \caption{Decay of metastable state in the presence of external field $h_{ext}= -0.120$ (coalescence regime): The variation of magnetisation with time (MCSS) in (a) the cubic shell of lattice size $L=40$; temperature maintained at $T = 1.82$ ($ 0.8 T_c^p$) and (b) the 3D solid cubic lattice of same size $L=40$; temperature fixed at $T = 3.608$ ($ 0.8 T_c$).}
    \label{fig:decay}
\end{figure}

To study the decay of the metastable state in the Ising ferromagnetic cubic shell we have initially considered a cubic shell of size $L=40$ with all Ising spins up, kept at a fixed temperature $T=1.82=0.8T_c^p$ below pseudo-critical temperature. Now the system is kept in an external magnetic field, say $h_{ext} = -0.12$, applied in the opposite direction of that of initial magnetisation. The system is observed to keep itself in a metastable state of positive magnetisation for a certain time duration, even in the presence of a negative external magnetic field and finally decay to the stable state of negative magnetisation, as shown in figure \ref{fig:decay}(a). It may be noted here, that multiple metastable states
(multiple step like variation of magnetisation) are found in the case of cubic shell. This has not been observed
\cite{acharyya1} before in the context of
two-dimensional Ising ferromagnet. This is an additional indication that the Ising ferromagnetic cubic shell behaves 
differently from that of a two-dimensional Ising ferromagnet.

\begin{figure}[h!tpb]
    \centering
    \includegraphics[angle=0, width=1.00\textwidth]{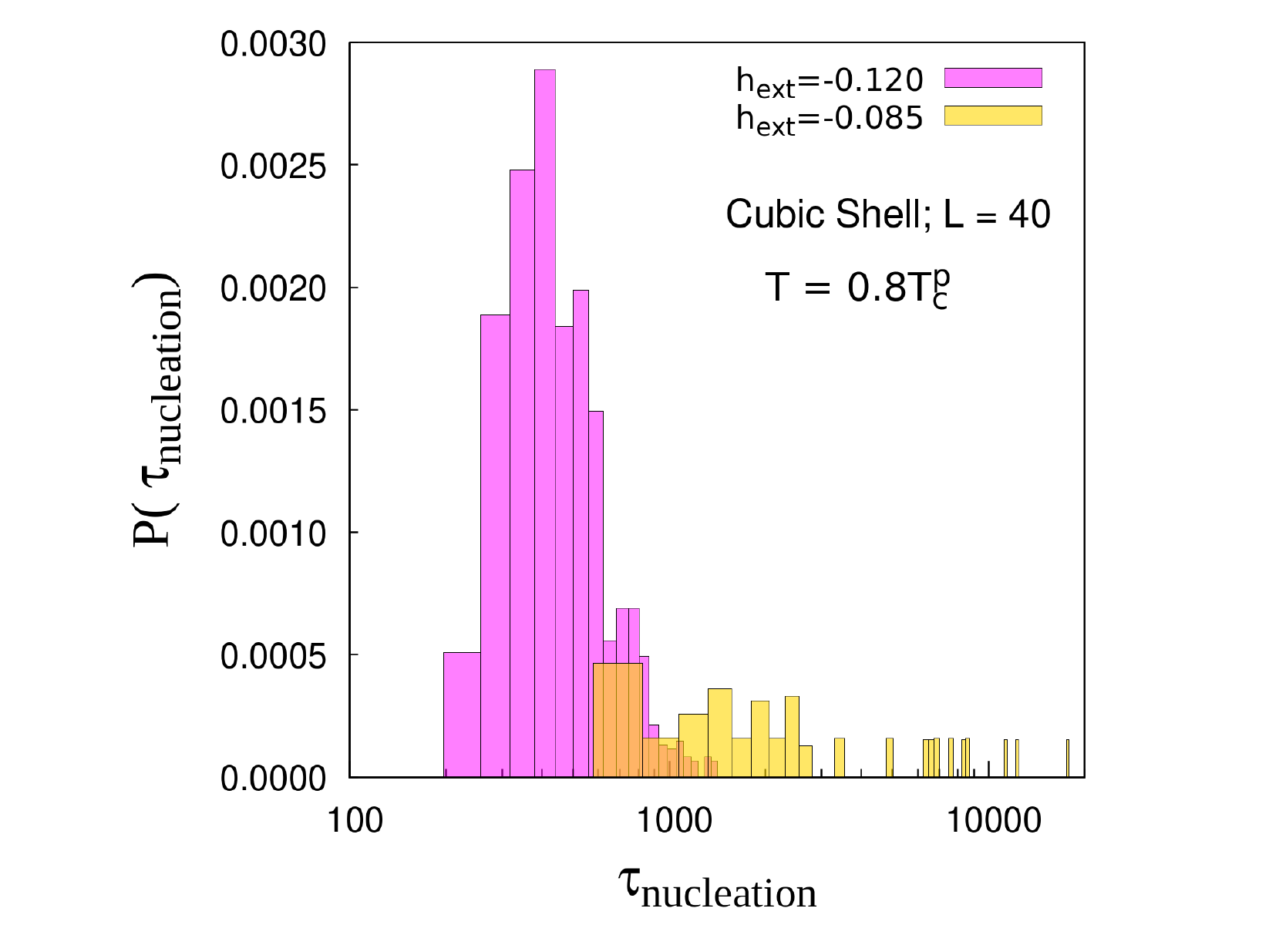}
    \caption{Normalised distribution of nucleation times ($\rm \tau_{nucleation}$) in coalescence regime (CR, $h_{ext} = -0.120$) and in nucleation regime (NR, $h_{ext} = -0.085$) for cubic shell lattice of size $L=40$ at temperature $T = 1.82$ ($0.8T_c^p$). The results are obtained for 1000 random samples.}
    \label{fig:distribution}
\end{figure}

The comparative study of metastable behaviours of Ising ferromagnetic cubic shell and solid cubic ferromagnetic system is demonstrated in figure \ref{fig:decay}. We have defined the metastable lifetime  as the minimum number of iterations
(MCSS) required to achieve the state below the cutoff magnetisation $m = 0.7$ \cite{rikvold}. Here the cutoff value has been chosen arbitrarily \cite{acharyya1, Stauffer}. In our study, the metastable lifetimes (or nucleation time)  are recorded for 1000 random samples at a fixed value of the external field. We investigated the statistical distribution of these metastable 
lifetimes. Fig-\ref{fig:distribution} shows the statistical distribution of the metastable lifetimes for $h_{ext} = -0.120$ (Pink) and $-0.085$ (Yellow) of an Ising ferromagnetic cubic shell of size $L=40$. The distribution graph indicates the enormous fluctuation in metastable lifetimes, leading to a considerably wider distribution for the field $h_{ext}=-0.085$ compared to 
that for the field $h_{ext}=-0.120$.

\begin{figure}[h!tpb]
\begin{center}
\includegraphics[angle=0, width=1.00\textwidth]{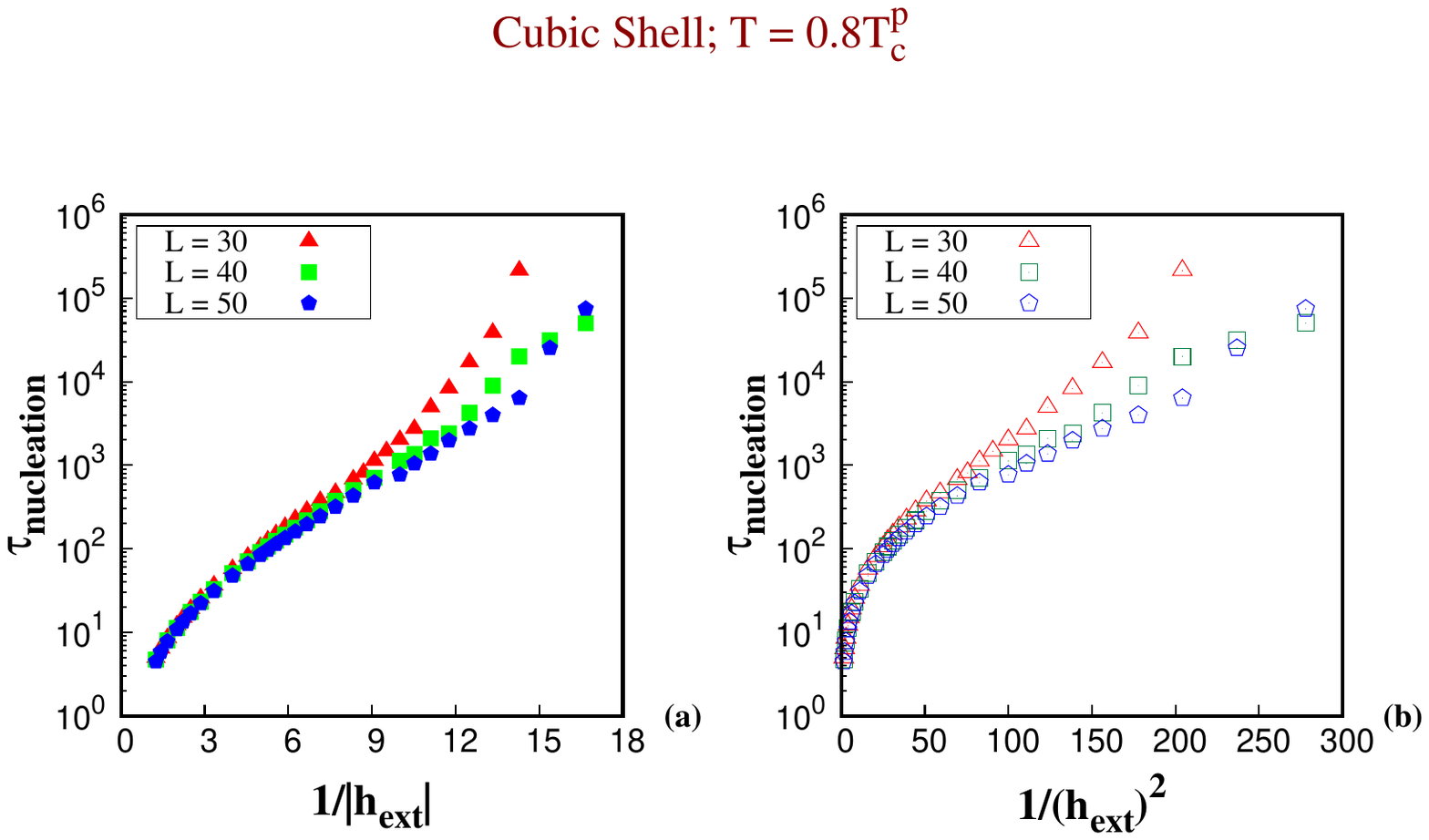}
\caption{(a) Nucleation time ($\rm \tau_{nucleation}$) as a function of $\rm 1/|h_{ext}|$ plotted in semi-logarithmic scale for Ising ferromagnetic cubic shell at fixed temperature $T = 1.82$ ($0.8T_c^p$). Different symbols indicate the data for different lattice sizes $L = 30$ (red solid triangle), $L = 40$ (green solid box) and $L = 50$ (blue solid pentagon). (b) Nucleation time ($\rm \tau_{nucleation}$) plotted as a function of $1/h_{ext}^2$ in the semi-logarithmic scale. Data obtained for ferromagnetic Ising ferromagnetic cubic shell  at fixed temperature $T = 1.82$ ($0.8T_c^p$). Different symbols indicate the data for different lattice sizes $L = 30$ (red hollow triangle), $L = 40$ (green hollow box) and $L = 50$ (blue hollow pentagon).} 
\label{fig:nuclntime-h1-h2}
\end{center}
\end{figure}

In our simulation, $\tau_{nucleation}$ is computed as the mean value of nucleation time (or metastable lifetime) to achieve $m=0.7$ across 1000 random samples in the strong field regime (SFR) and coalescence regimes (CR) where fluctuations are comparatively small. Conversely, in the nucleation regime (NR) or single droplet regime, where fluctuations are enormous we have considered the median value of nucleation times instead of the mean. This prescription has been followed by many related previous studies\cite{acharyya1}. Now, we have examined the nucleation time $\tau_{nucleation}$ as a function of the external field $h_{ext}$ for different sizes of cubic shell ($L=30$, 40 and 50) to analyze the impact of the external field on nucleation
(or metastability). The nucleation time $\tau_{nucleation}$ is plotted against $1/|h_{ext}|$ and $1/h_{ext}^2$ in semi-logarithmic scale represented in figure \ref{fig:nuclntime-h1-h2} (a) and (b) respectively. In these graphical plots three distinct regimes (i) strong field regime (SFR), (ii) coalescence regime (CR) and (iii) nucleation regime (NR) are definitively characterized. It is quite noticeable that the metastable lifetime (or nucleation time) remains almost independent of system size in both strong field regime (SFR)  and  coalescence regime (CR). On the other hand, at weaker field 
namely nucleation regime(NR), the nucleation time $\tau_{nucleation}$ decreases with increasing the system size($L$)
(as reported in \cite{acharyya1} for two- and three-dimensional Ising ferromagnetic system) for same amount of applied magnetic field $h_{ext}$. 

Now a question arises: why does nucleation in the Ising ferromagnetic cubic shell capture our interest?

\begin{figure}
    \centering
    \includegraphics[angle=0, width=1.00\textwidth]{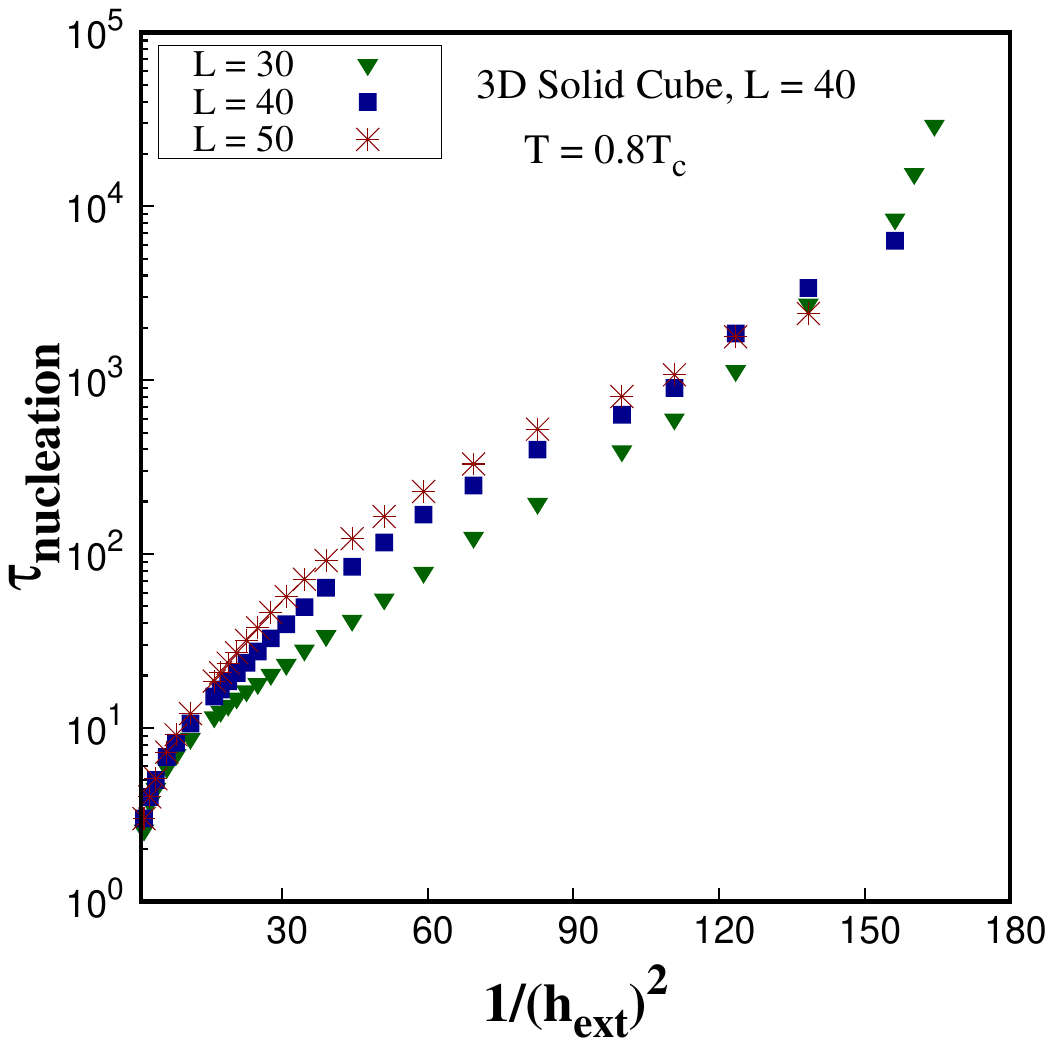}
    \caption{Nucleation time ($\rm \tau_{nucleation}$) as a function of $1/{h_{ext}^2}$ plotted in semi-logarithmic scale for three-dimensional ferromagnetic Ising model on a solid cubic lattice at fixed temperature $T = 3.608..$($0.8T_c$). Different symbols indicate the data for different lattice sizes $L = 30$ (green inverted triangle $\blacktriangledown$), $L = 40$ (blue box $\blacksquare$) and $L = 50$ (red asterisk $\boldsymbol{\ast}$).}
    \label{fig:nuclntime-solid}
\end{figure}

The nucleation process significantly depends on the structure, dimensionality and constraints of the lattice. In a three dimesional solid cubic lattice, each spin, inside the bulk, interacts with six nearest neighbours, while each spin at the surface (excluding the edge) can interact with five interacting spins. In contrast, spin in the cubic shell can experience the interaction with a maximum of four nearest neighbours. Consequently, the metastable behaviour of the cubic shell should differ from that of a three-dimensional solid cubic. Notably, the nucleation regime of a cubic shell initiates at a weaker field compared to that of the solid cube. The results obtained for the cubic shell are shown in figure \ref{fig:nuclntime-h1-h2}(b) while Fig-\ref{fig:nuclntime-solid} shows the results for the solid cube. \par

According to Classical Nucleation Theory (CNT), the metastable lifetime (or nucleation time) ($\tau $, which is inversely proportional to nucleation rate $I$) in the single-droplet or nucleation regime is formulated as \cite{becker} \[ \tau_{nr} \sim I^{-1} \sim \exp \Bigg(\frac{K_d \sigma^d} {k_BTh_{ext}^{d-1}} \Bigg) \]
Where `$d$' denotes the system's {\it Euclidean dimensionality}, $\sigma$ stands for surface tension and $K_d$ is a $d$ dependant constant. In the coalescence regime, multiple supercritical droplets form and grow simultaneously, they coalesce 
to form bigger droplets and finally engulf the whole system. In the multi-droplet or coalescence regime, the 
metastable lifetime (or nucleation time) is given by, \cite{becker} \[ \tau_{cr} \sim I^{- \frac{1}{(d+1)}} \sim \exp \Bigg(\frac{K_d \sigma^d} {(d+1) k_BTh_{ext}^{d-1}} \Bigg) \] Therefore we may conclude that the logarithm of 
the metastable lifetime (or nucleation time) will follow linear relation with $\frac{1}{h_{ext}^{d-1}}$; i.e., $\log (\tau_{nr}) \sim \frac{1}{h_{ext}^{d-1}}$ and $\log (\tau_{cr}) \sim \frac{1}{(d+1) h_{ext}^{d-1}}$. These theoretical prediction has been confirmed by large scale Monte Carlo simulation using multispin coding algorithm\cite{acharyya1}. The ratio of the slope in the nucleation regime(NR) to that in the coalescence regime(CR) is denoted as $R=\frac{\rm Slope~in~NR}{\rm Slope~in~CR} = d+1$. 

\begin{figure}[h!tpb]
\begin{center}
\includegraphics[angle=0, width=1.00\textwidth]{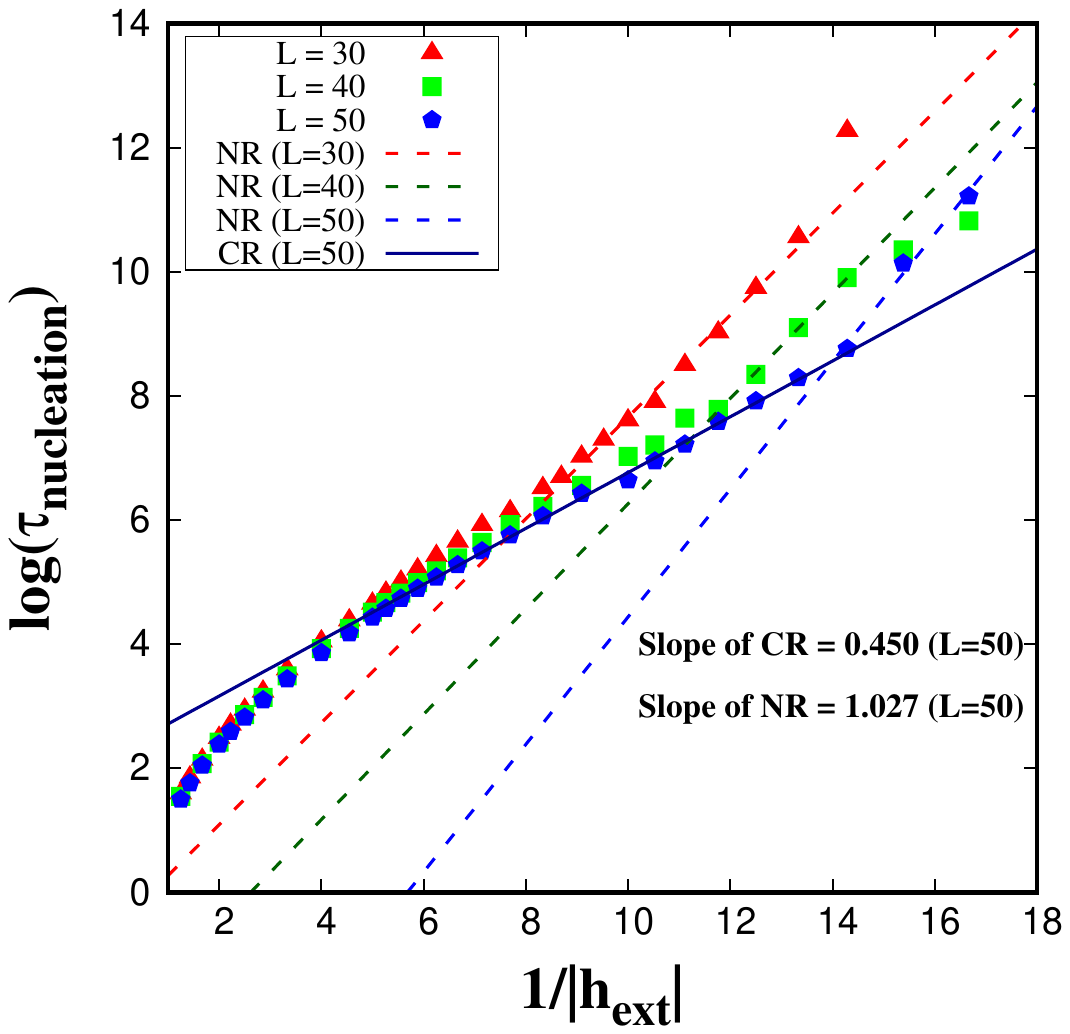}
\caption{$\rm \log (\tau_{nucleation})$ is plotted against $1/|h_{ext}|$ for different system sizes $L=30, 40~\rm{and}~50$ at temperature $T=0.8T_c^p$ for Ising ferromagnetic cubic shell. The dashed lines represent the linear best fit in the nucleation regime (NR) and the solid line indicates the best-fitted line in the coalescence regime (CR). The slope of the best-fitted line in the nucleation regime = 1.027 (blue dashed line) and the slope of the best-fitted line in the coalescence regime = 0.450 (blue solid line) for lattice size $L=50$.}
\label{fig:nuclntime-h1}
\end{center}
\end{figure}

\begin{figure}
\includegraphics[angle=0, width=1.00\textwidth, center]{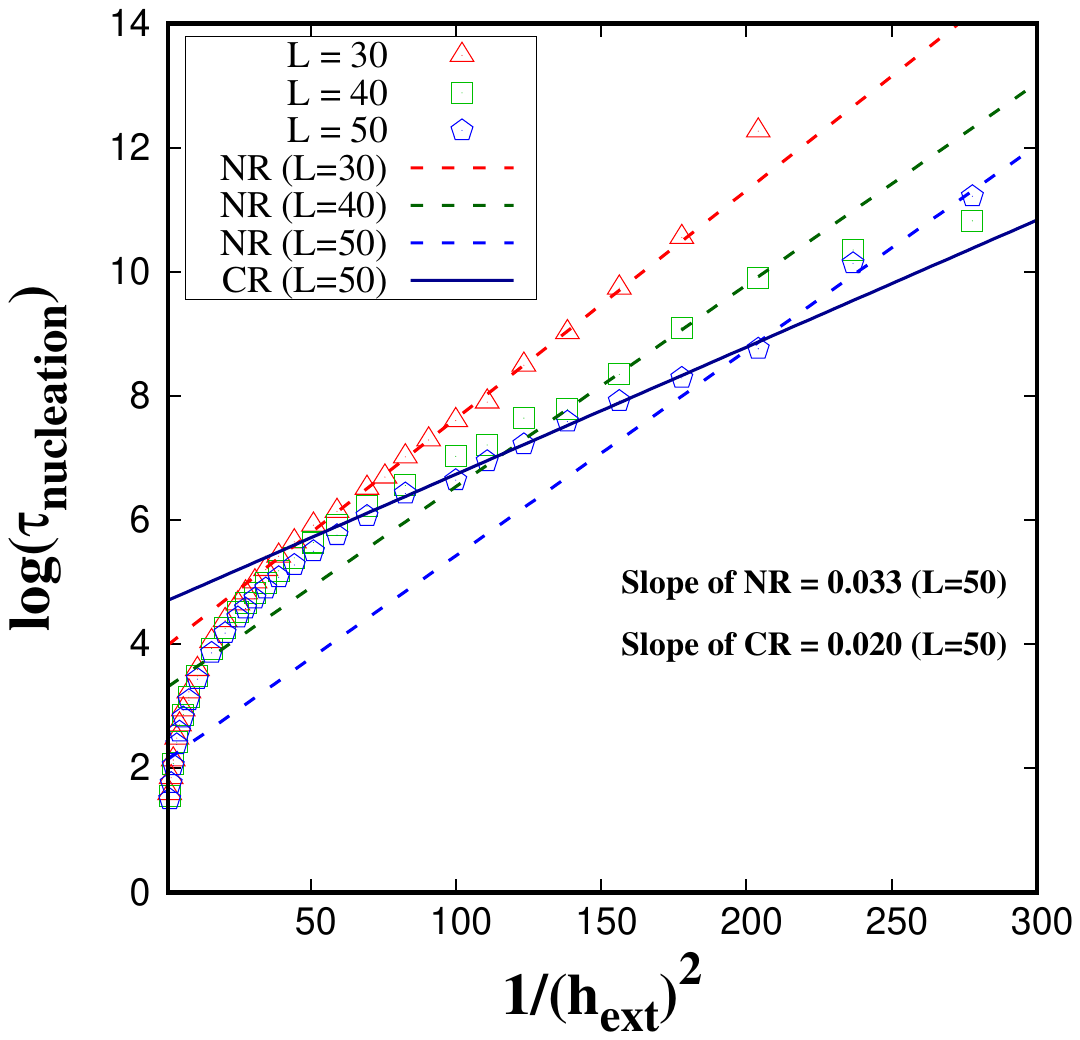}
\caption{$\rm \log (\tau_{nucleation})$ is plotted against $1/h_{ext}^2$ for different system sizes $L=30, 40~\rm{and}~50$ at temperature $T=0.8T_c^p$ for Ising ferromagnetic cubic shell. The dashed lines represent the linear best fit in the nucleation regime (NR) and the solid line is for the coalescence regime (CR). The slope of the best-fitted line in the nucleation regime = 0.033 and the slope of the best-fitted line in the coalescence regime = 0.020 (for lattice size $L=50$).}
\label {fig:nuclntime-h2}
\end{figure}

In figure \ref{fig:nuclntime-h1} the logarithmic value of nucleation time ($\log(\tau_{nucleation})$) is plotted against the inverse of external field strength ($1/|h_{ext}|$) for the cubic shell  of size $L=30$, 40 and 50. This plot exhibits the linear relation between $\log(\tau_{nucleation})$ and $1/|h_{ext}|$ with distinct slopes characterizing the coalescence (or multi-droplet) region and the nucleation (or single-droplet) region. The solid line indicates the best-fitted line for the coalescence regime and the dashed lines give the slope for the nucleation regime. We may conclude that our observed results, depicted in figure \ref{fig:nuclntime-h1} are consistent with the Classical Nucleation theory. The slopes in both nucleation and coalescence regions are determined from linear best fit and represented in tabular form of Table \ref{tab:table1}.
\begin{table*}[h!]
  \caption {Best fitted parameters (Cubic shell) corresponding to figure \ref{fig:nuclntime-h1}.}
  \label{tab:table1}
  \begin{tabular*}{\textwidth}{@{\extracolsep{\fill}}lllll}
    \hline
    Region & Size (L) & Slope (a)  &   $\chi^2$ &  DOF  \\ 
    \hline 
    Coalescence regime & 40 & $0.498 \pm 0.005$ & 0.003 & 7 \\
        & 50 & $0.450 \pm 0.005 $ & 0.032 & 12  \\
        
    \hline    
    Nucleation regime & 30 & $ 0.821 \pm 0.027 $  & 0.100 & 7  \\
        & 40 & $0.848 \pm 0.016 $ & 0.002 & 2    \\
       
        & 50 & $1.028 \pm 0.118$ & 0.039 & 1  \\  
    \hline
  \end{tabular*}
\end{table*}
The theoretically predicted value of slope-ratio in two-dimension $d=2$ is, $R_{th}\Big|_{d=2}=d+1=3$.
In our simulation for Ising ferromagnetic cubic shell, the estimated value of $R$; $R_{exp}\Big|_{d=2}=2.284$ for lattice size $L=50$  and  $R_{exp}\Big|_{d=2}=1.703$ for $L=40$. The percentage of error  $=\Big(\frac{|R_{exp} - R_{th}|}{R_{th}}\Big) \% \bigg|_{d=2}= 23.9 \%$; for $L=50$ and $43.2 \%$; for $L=40$.\\
We have also plotted the logarithmic of nucleation time $\log(\tau_{nucleation})$ as the function of $1/h_{ext}^2$ in figure \ref{fig:nuclntime-h2}, exhibiting three distinct regions (SFR, CR and NR) and estimated the slopes in Coalescence (marked by solid line) and Nucleation (indicated by dashed lines) regimes, as represented in table \ref{tab:table2}.
\begin{table*}[h!]
  \caption {Best fitted parameters (Cubic shell) corresponding to figure \ref{fig:nuclntime-h2}.}
  \label{tab:table2}
  \begin{tabular*}{\textwidth}{@{\extracolsep{\fill}}lllll}
    \hline
    Region & Size (L) & Slope (a)  &   $\chi^2$ &  DOF  \\ 
    \hline 
    Coalescence regime & 40 & $0.0244 \pm 0.0013$ & 0.003 & 2 \\
        & 50 & $0.0205 \pm 0.0007 $ & 0.018 & 5  \\
        
    \hline    
    Nucleation regime & 30 & $0.0368 \pm 0.0004 $  & 0.032 & 10  \\
        & 40 & $0.0325 \pm 0.0006 $ & 0.002 & 2    \\
       
        & 50 & $0.0331 \pm 0.0045$ & 0.056 & 1  \\  
    \hline
  \end{tabular*}
\end{table*}
 The theoretically predicted value of slope-ratio $R$ for the dimension $d=3$ is $R_{th}\Big|_{d=3}= d+1=4$. We have determined the value of slope-ratio $R$ in the case of $\log(\tau_{nucleation})\sim 1/h_{ext}^2$; $R_{exp}\Big|_{d=3}=1.615$ of the cubic shell of size $L=50$  and  $R_{exp}\Big|_{d=3}=1.332$ for $L=40$. The percentage of error  $=\Big( \frac{|R_{exp} - R_{th}|}{R_{th}}\Big) \% = 59.6 \%$; for $L=50$ and $66.7 \%$; for $L=40$. \\
 
 Therefore the statistical analysis confirms that the metastable behaviour of an Ising ferromagnetic cubic shell is closer to $d=2$ rather than $d=3$ in the context of the Classical Nucleation theory. According to Euclidean geometry, the dimension of a cubic shell or hollow cube is three. Although our simulation predicts that the nucleation of metastable states in the Ising ferromagnetic cubic shell mostly obeys the prediction of classical nucleation theory for two-dimensional system. 
 
\begin{figure}
    \centering
    \includegraphics[width=1.00\textwidth]{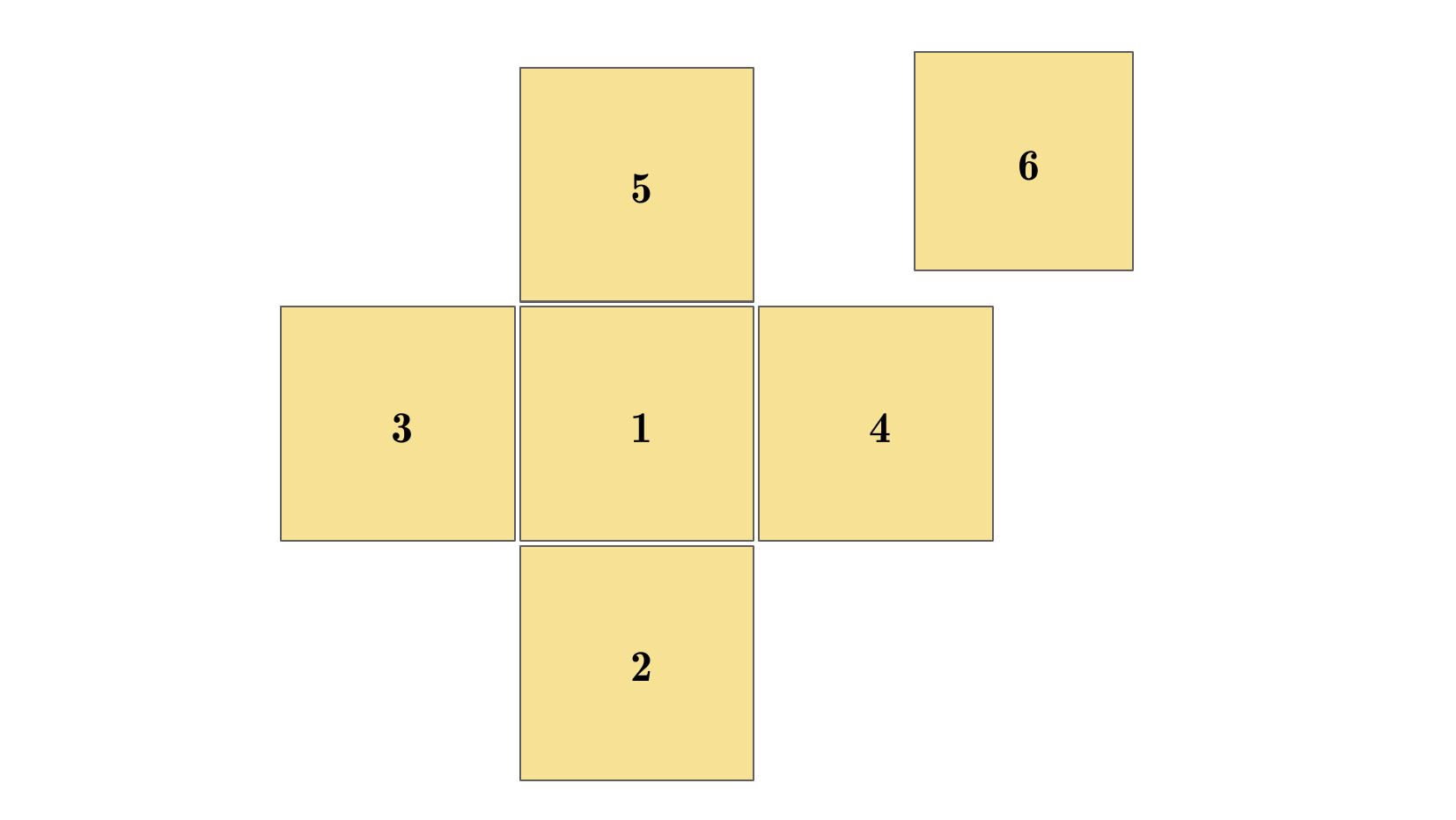}
    \caption{Demonstration of the unfolded cubic shell, following the standard dice layout for face numbering.}
    \label{fig:surface}
\end{figure}

\begin{figure}
\centering
  \includegraphics[angle=0,width=0.465\textwidth]{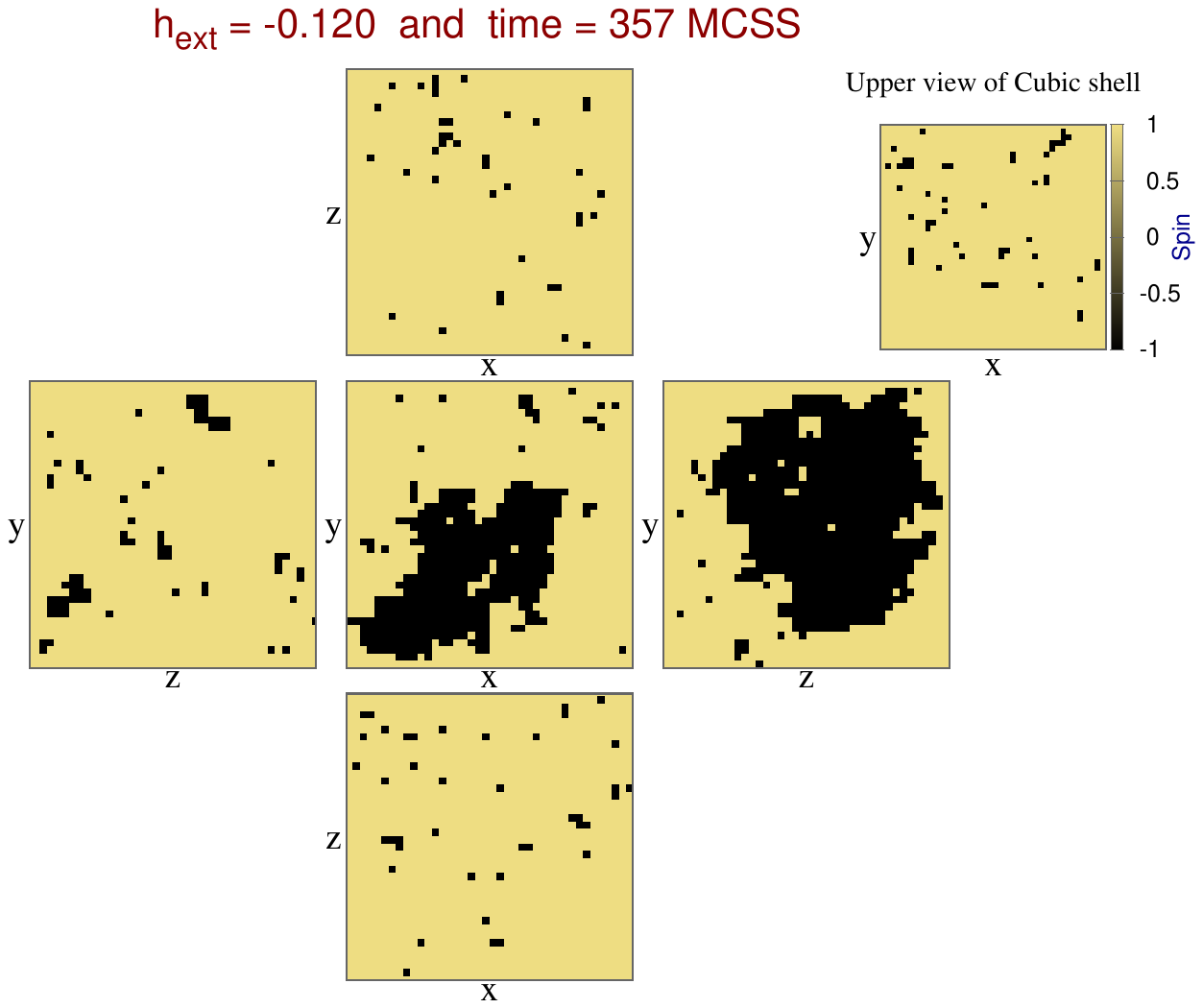}(a)
  \includegraphics[angle=0,width=0.466\textwidth]{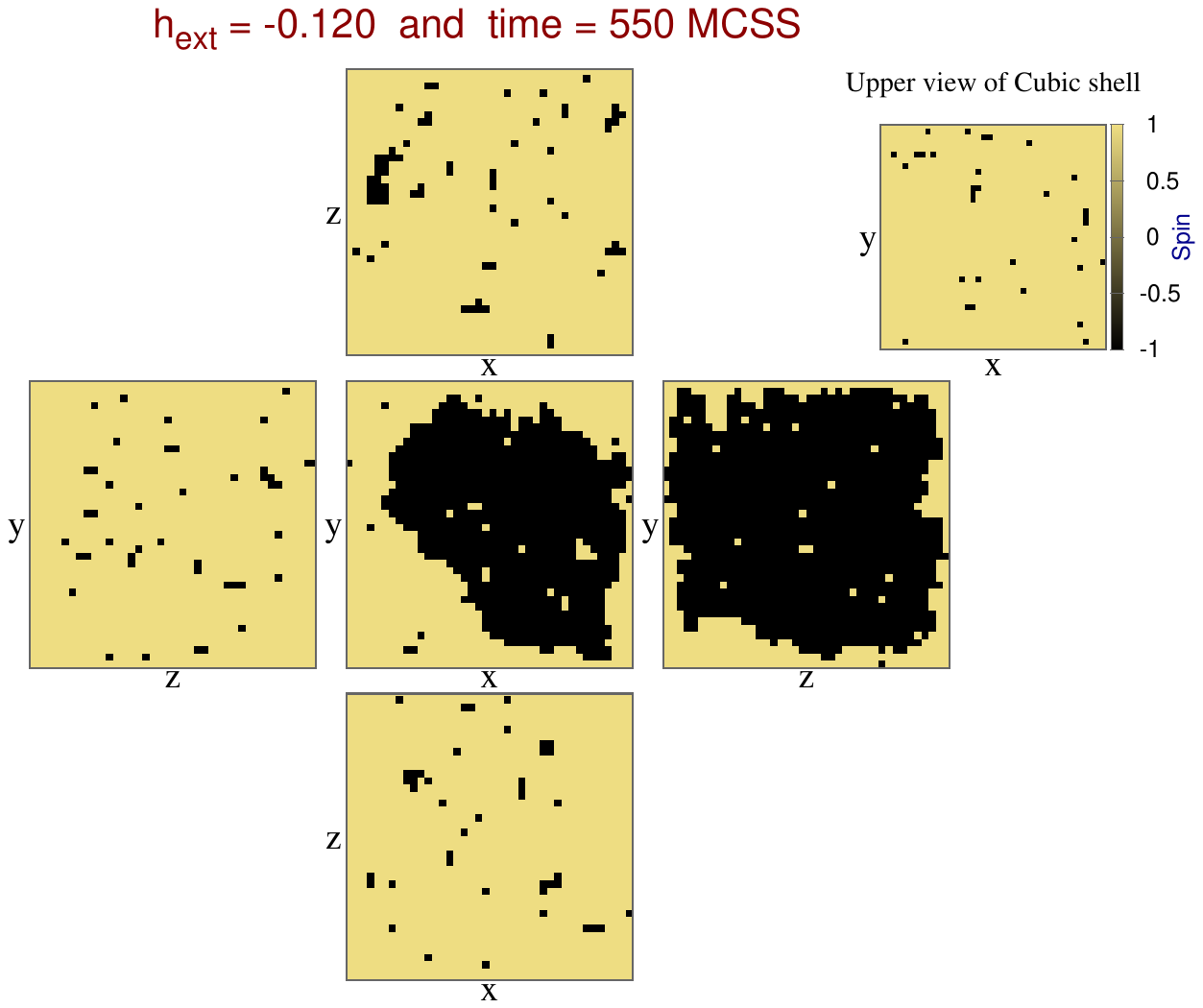}(b)\\
  \includegraphics[angle=0,width=0.90\textwidth]{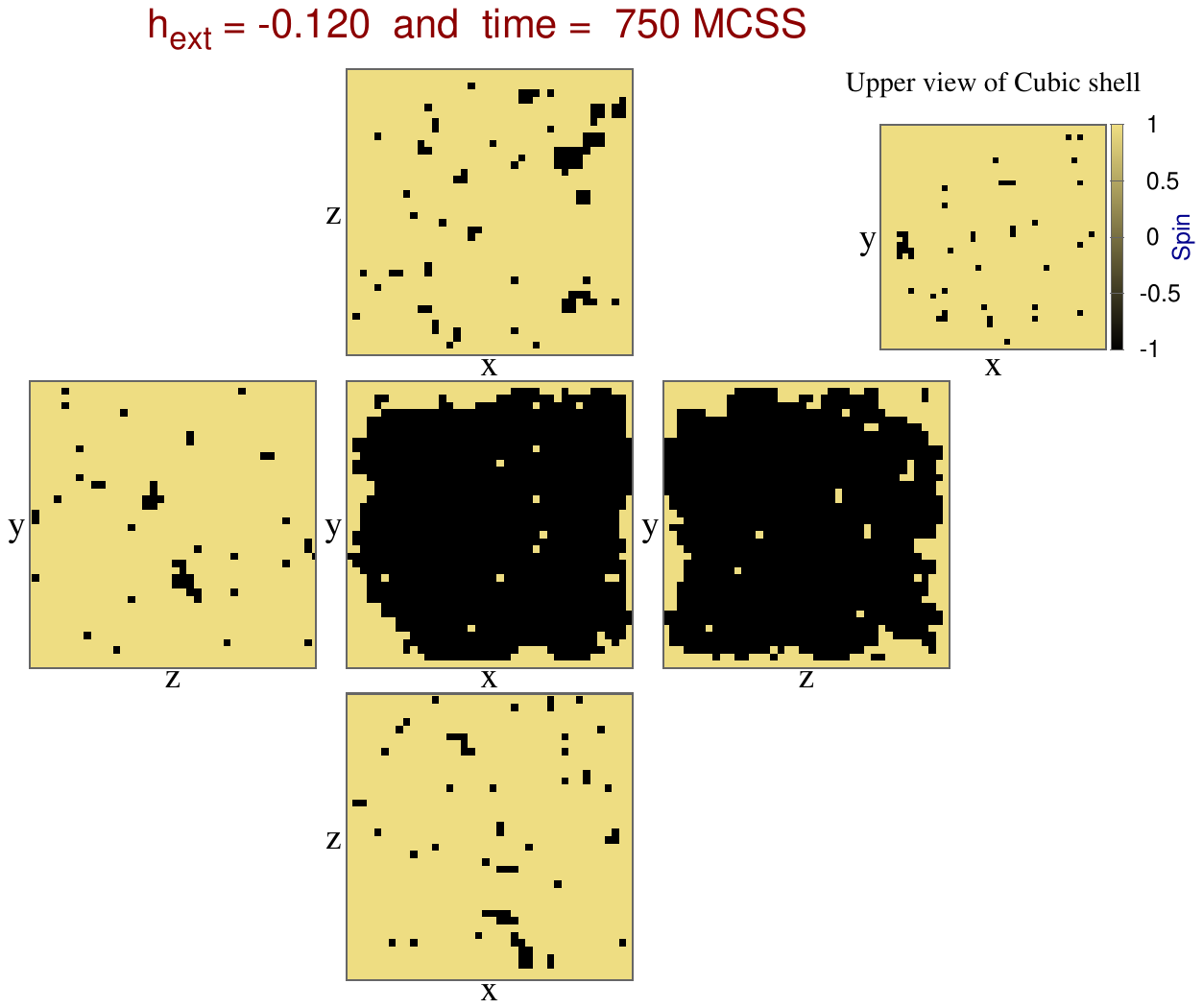}(c)
  \caption{Image plots of the spin configuration ($L=40$) at three different time instants (in unit of MCSS) in the presence of external field $h_{ext}=-0.120$ i.e., coalescence regime. (a) $t = 357$ (after reaching the state of $m=0.7$), (b) $t = 550$ and (c) $t = 750$. }
  \label{fig:snapshot-CR} 
\end{figure}

\begin{figure}
  \centering
  \includegraphics[angle=0, width=0.465\textwidth]{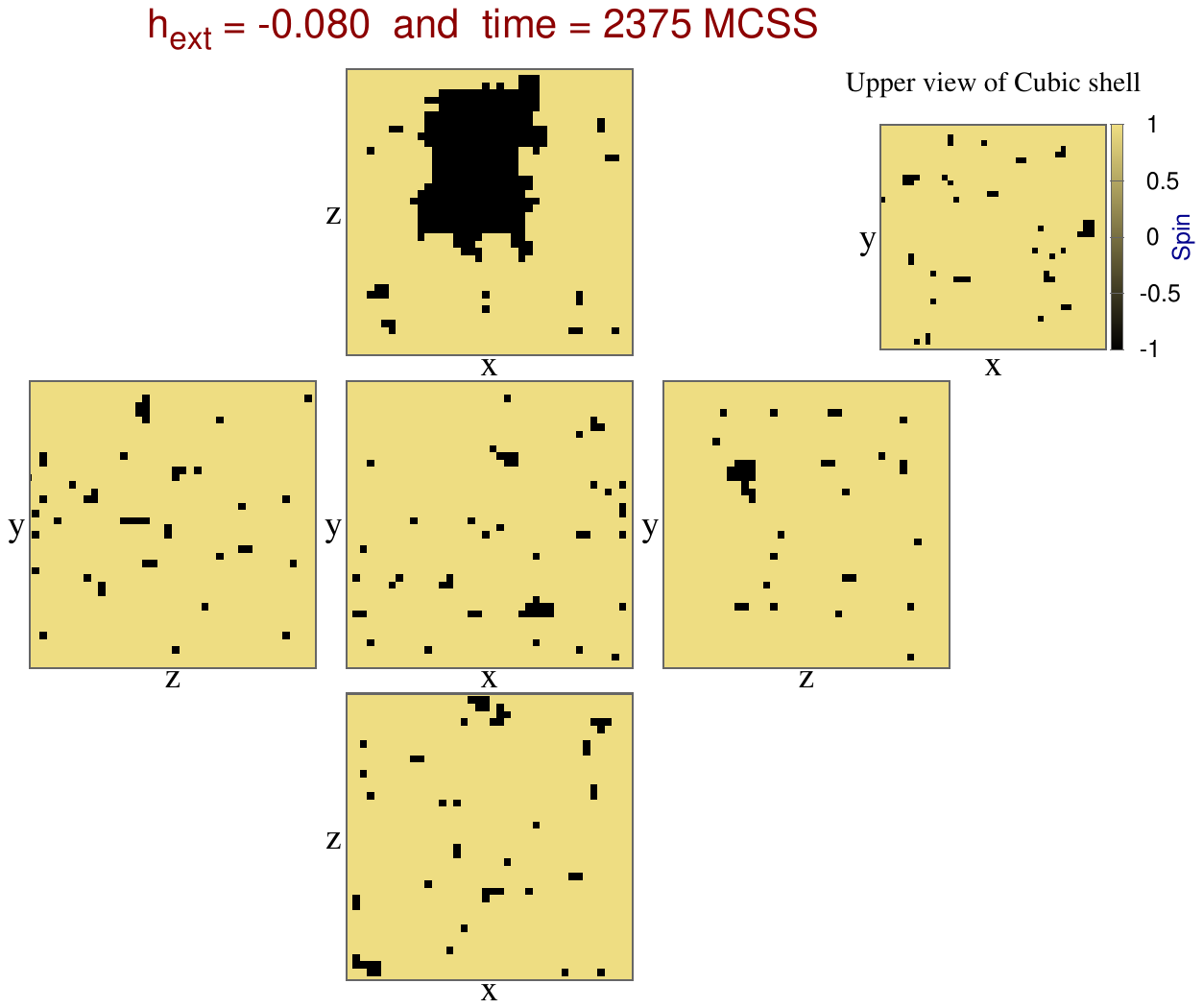}(a)
  \includegraphics[angle=0, width=0.466\textwidth]{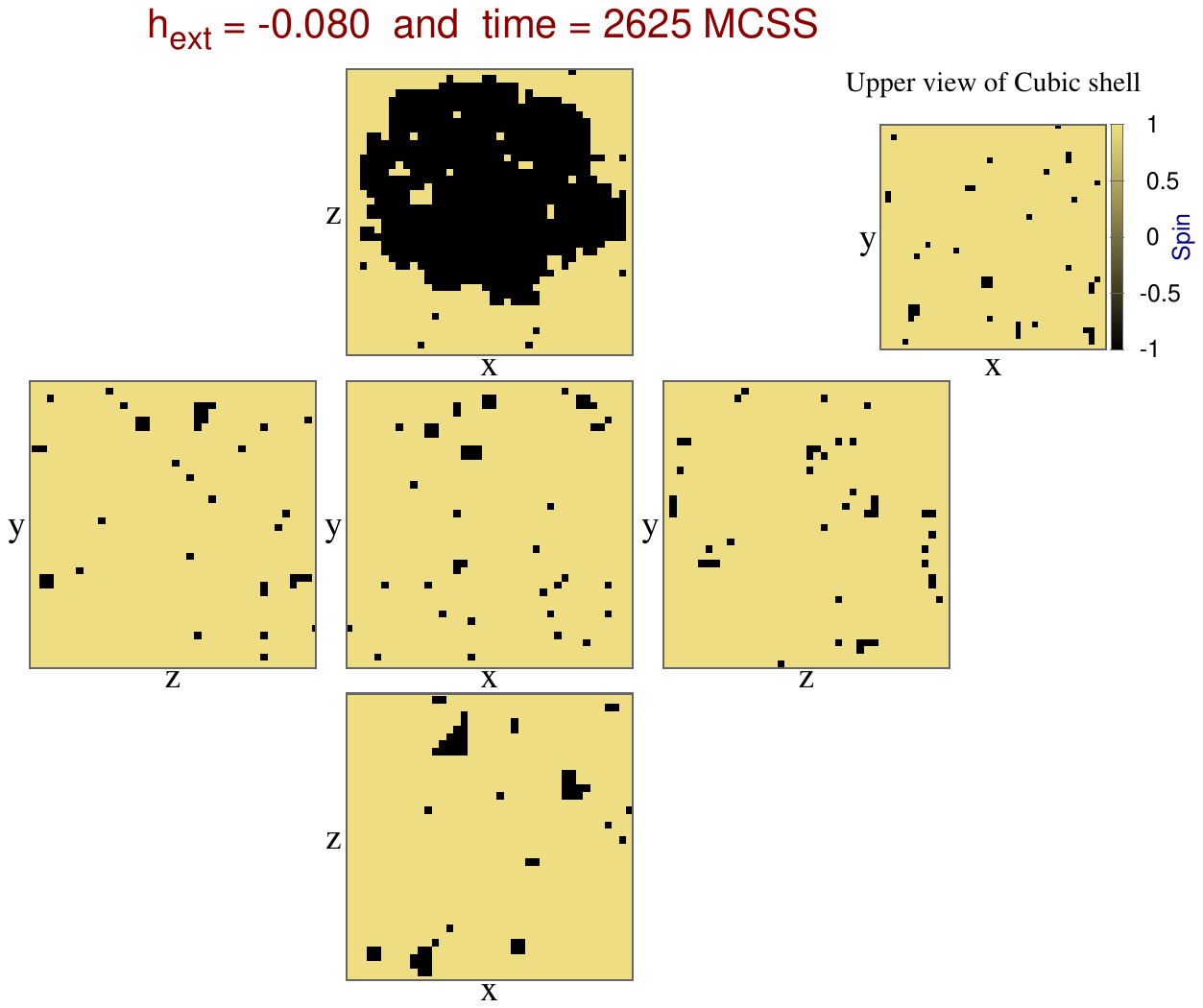}(b)\\
  \includegraphics[angle=0, width=0.90\textwidth]{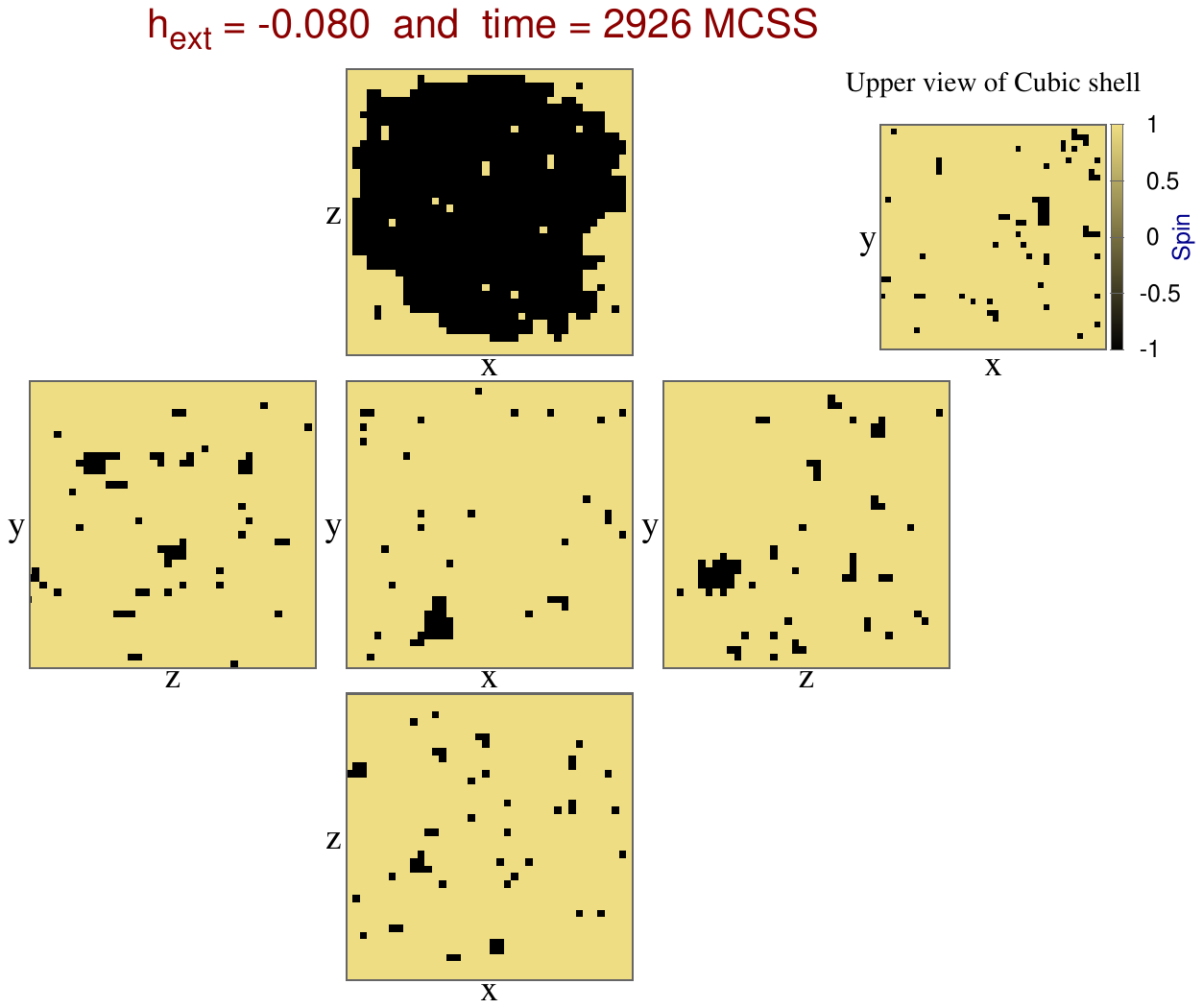}(c)
  \caption{Snapshots of the spin configuration ($L=40$) at three different time instants (in unit of MCSS) in the presence of external field $h_{ext}=-0.080$ i.e., nucleation regime. (a) $t=2375$, (b) $t=2625$ and (c) $t=2926$ (after reaching the state of $m=0.7$). }
  \label{fig:snapshot-NR} 
\end{figure}

 As a complementary, we have studied the growth of droplets in the metastable behaviour of Ising ferromagnetic cubic shell.
 We have taken the snapshots of spin-configurations of various planes of  the cubic shell  ($L=40$) at three different times (MCSS unit) and presented in figure \ref{fig:snapshot-CR}  and \ref{fig:snapshot-NR} for coalescence and nucleation regime respectively. The snapshots illustrate the droplet theory of nucleation. In figure \ref{fig:surface} we have shown the unfolded view of a cubic shell following the numbering convention of a standard dice of Ludo. The faces are numbered as follows: The XY plane at $Z=0$ is designated as face 1 and the XY plane $Z=L$ is defined as face 6. The XZ plane at $Y=0$ is labeled as face 2 and $Y=L$ as face 5. Similarly, the YZ plane at $X=0$ is labeled as face 3 and $X=L$ as face 4. The colour codes used in the image plot are: Yellow colour represents the up or $S_i=+1$ spins and Black indicates the down spin $S_i=-1$. 
 
 In the coalescence regime (e.g $h_{ext} = -0.12$) a pair of supercritical droplets (connected domain) of down spin form and evolve simultaneously on the faces 1 and 4 of the cubic shell  and after a certain time iterations these two droplets merge along the edge boundary of the lattice, creating a larger domain of negative spins, as shown in figure \ref{fig:snapshot-CR}. Figure \ref{fig:snapshot-CR} (a) illustrates how two clusters are created at the time after the system achieves the state of $m=0.7$. Subsequently, another two droplets of down spin, larger than critical size nucleate on face 2 and face 3 around the same time and gradually coalesce with other droplets. This process continues until the entire system (all planes) is nearly engulfed by the droplets or clusters of negative spins. The nucleation phenomena in the Ising cubic shell at Coalescence regime are clearly explained by the timelapse video of growing and shrinking droplets, found on YouTube with the link \href{https://youtu.be/7AeoOowC07g}{\underline{https://youtu.be/7AeoOowC07g}}. Whereas in the nucleation regime (e.g. $h_{ext} = -0.08$), a single droplet of supercritical size  on face 5  grows to occupy the majority of that face, as demonstrated in figure \ref{fig:snapshot-NR}. A similar phenomenon successively occurs on the other faces of the cubic shell  (e.g. 3, 4, 2 and 1 etc.). The single droplet of negative spin forms at a time on each face one after another and then gradually evolves to invade the system, as explained by the timelapse video of growing  droplets, found on YouTube with the link \href{https://youtu.be/QyNjoeC8rSg}{\underline{https://youtu.be/QyNjoeC8rSg}}. It is worth mentioning that a single droplet (cluster) developed on one of the six faces of the cubical shell does not penetrate to other faces which confirms the growth in the nucleation regime.

\section {\bf {Summary:}}
 The equilibrium and nonequilibrium properties of an Ising ferromagnetic cubic shell have been studied extensively by Monte Carlo simulation. The results are compared with that of a solid cube. The cubical shell differs from a solid cube by the number of nearest neighbours of the spins on the surface. Geometrically, the Euclidean dimensions of both the solid cube and the cubical shell are three. However, it is noticed from detailed Monte Carlo analysis that the Ising ferromagnetic cubical shell behaves like a two-dimensional object in the context of ferro-para phase transition. Our results of the equilibrium magnetisation as a function of the temperature of a cubical shell show ferromagnetic phase transition near a pseudocritical temperature which is close to that for the two-dimensional Ising ferromagnet. That is an interesting result, we believe. The transition temperature of the cubical shell has been found to be the temperature which maximizes the
susceptibility. The pseudocritical temperature $T^p_c \simeq 2.28$ is close to the value of that obtained from the exact calculation by Onsager\cite{Onsager}.

The surprising results are obtained in studying the relaxation behaviour of the Ising ferromagnetic cubic shell. 
Expecting the exponential relaxation (as mentioned in earlier studies of a two dimensional Ising ferromagnet), we got surprised when we have observed the
non-exponential relaxation of cubic shell. Moreover, our results show that the relaxation of an 
Ising ferromagnetic cubic shell is not even stretched exponential\cite{Stauffer}. In this case, the cubic shell does not behave like two-dimensional system as
predicted from the results of ferro-para phase transition of Ising ferromagnetic cubic shell.

We have also investigated the metastable behaviours of the cubical Ising ferromagnetic shell. The well-persistent 
{\it multiple} metastability has been observed in the weak field limit. The metastable lifetime has been calculated. This metastable lifetime has been studied as a function of the applied magnetic field. The three regions namely, the strong field regime, coalescence regime and nucleation regimes have been identified distinctly. These results are qualitatively consistent with the prediction
of Becker-D\"oring classical nucleation theory. However, the statistical analysis matches more likely with the two-dimensional metastable behaviour of Ising ferromagnet.

In conclusion, we can say that the Euclidean dimension of an Ising ferromagnetic cubic shell cannot be determined certainly,
in the case of its equilibrium and nonequilibrium magnetic properties. It would be interesting to study the variations of the
ferro-para transition temperature with the thickness of the cubic shell.
 \textcolor{blue}{This requires huge computational efforts and is beyond the scope of the present study. Moreover, the critical exponents for the magnetisation ($M(T_c) \sim L^{-{{\beta} \over {\nu}}}$) and susceptibility ($\chi_{max} \sim L^{{\gamma} \over {\nu}}$) can be estimated  
(through detail finite size analysis) to capture the effects of the Euclidean dimension ($d$) of the system. 
These will be considered in a separate project and the results will be reported elsewhere. 
The metastable behaviours of other two
dimensional systems (e.g., Kagome lattice) would  be an interesting study.} 

\vskip 0.5cm

\noindent {\bf Acknowledgements:}  
I. Tikader acknowledges UGC JRF, Govt. of India for financial support. 
\vskip 0.2cm


\noindent {\bf Data availability statement:} Data will be available on request to Ishita Tikader.

\vskip 0.2cm

\noindent {\bf Conflict of interest statement:} We declare that this manuscript is free from any conflict of interest.
\vskip 0.2cm

\noindent {\bf Funding statement:} No funding was received particularly to support this work.

\vskip 0.2cm

\noindent {\bf Authors’ contributions:} Ishita Tikader- Developed the code, prepared the figures, and wrote the manuscript.
Muktish Acharyya-Conceptualized the problem, analysed the results and wrote the manuscript.


\newpage

\newpage
\newpage
\newpage
\newpage
\newpage
\newpage

\newpage
\newpage
\newpage
\newpage

\newpage
\newpage

\end{document}